\documentclass[a4paper,11pt,twoside]{article}
\pdfoutput=1

\usepackage{geometry} 
\usepackage{a4wide}

\usepackage{amssymb,amsmath,bm}
\usepackage{cite}
\usepackage{url}
\usepackage{color}
\usepackage{slashed}
\usepackage{graphicx}
\usepackage[latin1]{inputenc}
\usepackage{amsfonts}
\usepackage{lscape}
\usepackage{amsthm}
\usepackage{booktabs}
\usepackage{array}
\usepackage{rotating}
\usepackage[numbers, sort]{natbib}
\usepackage{float}
\usepackage{xspace}

\usepackage{mathrsfs}

\usepackage[small,bf]{caption}
\selectfont

\begin{document}

\begin{titlepage}

\vspace*{-15mm}
\begin{flushright}
FTPI-MINN-16-09
\end{flushright}
\vspace*{0.7cm}

\begin{center}
{
\bf\huge
Improving naturalness in Gauge Mediation\\
\vspace{11pt}
with non-unified messenger sectors
}
\\[10mm]
{\Large Lorenzo~Calibbi$^{\,\star}$} \footnote{E-mail: \texttt{calibbi@itp.ac.cn}},
{\Large Tianjun~Li$^{\,\star\,\dagger}$} \footnote{E-mail: \texttt{tli@itp.ac.cn}},
{\Large Azar~Mustafayev$^{\,\ddag}$} \footnote{E-mail: \texttt{mustafayev@physics.umn.edu}},
{\Large Shabbar~Raza$^{\,\star}$} \footnote{E-mail: \texttt{shabbar@itp.ac.cn}}\\[1mm]
\end{center}
\vspace*{0.50cm}
\centerline{$^{\star}$ \it
State Key Laboratory of Theoretical Physics and Kavli Institute for Theoretical Physics}
\centerline{\it
China (KITPC), Institute of Theoretical Physics, Chinese Academy of Sciences,}
\centerline{\it Beijing 100190, P.~R.~China}
\vspace*{0.2cm}
\centerline{$^{\dagger}$ \it
School of Physical Electronics, University of Electronic Science and Technology of China,}
\centerline{\it Chengdu 610054, P.~R.~China}
\vspace*{0.2cm}
\centerline{$^{\ddag}$ \it
William I.~Fine Theoretical Physics Institute, University of Minnesota,}
\centerline{\it
Minneapolis, MN 55455, USA}
\vspace*{1.20cm}
\begin{abstract}
\noindent We study models of gauge-mediated supersymmetry breaking with messengers that do not belong to complete 
representations of grand-unified gauge groups. We show that certain setups characterized by heavy Wino
can greatly improve the fine tuning with respect to models with unified messengers, such as minimal gauge mediation.
The typical models with low tuning feature multi-TeV superparticles, with the exception of the Higgsinos and possibly Bino and right-handed sleptons. As a consequence, the absence of signals for supersymmetry at the LHC is trivially accommodated in our framework. On the other hand, testing these models will be challenging at the LHC. 
We finally show that the gravitino can be a consistent candidate for cold dark matter, 
 provided a rather low reheating temperature, if a standard thermal history of the universe is assumed.
\end{abstract}

\end{titlepage}

\addtocounter{footnote}{-4}

\section{Introduction}
\label{sec:intro}
Low-energy supersymmetry (SUSY) represents one of the most convincing frameworks to address to electroweak hierarchy
problem, as well as a common perturbative description of physics beyond the Standard Model (SM) from the electroweak scale up to very high energy scales, including gauge coupling unification and dark matter. 
Nevertheless, the discovery at the LHC of a Higgs scalar with mass about 125 GeV \cite{Aad:2012tfa,Chatrchyan:2012xdj}, 
as well as the negative results of a large number of dedicated SUSY searches, put the SUSY paradigm under stress,
leading to a severe fine tuning, at least within minimal SUSY models, see e.g.~\cite{Craig:2013cxa,Arvanitaki:2013yja}. 
This challenges the original motivation of SUSY and calls for a departure from minimality. 

In the present paper, we work within the context of the minimal supersymmetric standard model (MSSM) with
gauge-mediated supersymmetry breaking \cite{Giudice:1998bp}.
We are going to study how to relax the fine tuning of gauge mediation (GM) models, focussing on the possibility 
that the GM messengers belong to incomplete representations of grand-unified gauge groups. 
Such messenger sectors lead to a departure from the minimal models as they feature non-unification of gaugino masses.
We aim at identifying patterns of gaugino masses at the messenger scale that can lead to a reduced fine tuning, 
similarly to previous studies performed within the context of gravity mediation 
\cite{Gogoladze:2009bd,Horton:2009ed,Antusch:2012gv,Gogoladze:2012yf,Gogoladze:2013wva,Kaminska:2013mya,Martin:2013aha,Kowalska:2014hza}.
The results of this analysis can then give us a guideline on how to build models of messengers featuring a reduced fine tuning, along the lines of \cite{Li:2010hi}.
Specific choices of messengers and gaugino mass ratios leading to a reduced fine tuning in GM have been previously discussed in \cite{Brummer:2012zc,Brummer:2013yya,Bhattacharyya:2015vha,Fukuda:2015pra}.
Here we perform a full scan over the possible sets of non-unified messengers, and attempt a full characterization of
typical spectra and phenomenological consequences (especially for collider experiments) 
of the solutions featuring lowest fine tuning. 

Other directions leading to a reduced tuning or even natural SUSY scenarios (in few cases within the context of GM as well) 
have been recently explored in \cite{Li:2014xqa,Casas:2014eca,Li:2014dna,Du:2015una,Ding:2015vla,Ding:2015epa,Li:2016ucz,Casas:2016xnl}.

The rest of the paper is organized as follows. In section \ref{sec:model} we described our setup and specify the parameters 
we will study in the numerical analysis. In section \ref{sec:FT} we review the definitions of fine tuning we adopt in this study. The results of our numerical scan are shown and discussed in section \ref{sec:numerics}, while details of the spectra of models featuring low fine tuning, as well as some benchmark models, are presented in section \ref{sec:spectrum}. Collider constraints and prospects are discussed in section \ref{sec:collider}, while section \ref{sec:DM} is devoted to a discussion of the gravitino cosmology of the benchmark models. Finally, details about the model building aspects of GM
with non-unified messengers are given in section \ref{sec:models} and the outcome of our work is summarized in section
\ref{sec:conclusions}.
 
\section{Non-unified Gauge Mediation}
\label{sec:model}
In order to scan over models with non-unified messengers, 
we  adopt as free parameters the contribution of the GM messengers to the $\beta$-function coefficients of the SM ($SU(3)\times SU(2)\times U(1)$) gauge couplings, $g_1,~g_2,~g_3$:
\begin{equation*}
b_1^M,~b_2^M,~b_3^M.
\end{equation*}
These quantities are given by the sum of the Dynkin indices for the SM representations of the messengers.
For instance, each copy of $\mathbf{5+\bar{5}}$ corresponds to
$b_1^M=b_2^M=b_3^M=1$, as in minimal GM. Details about the messenger sector are discussed in section \ref{sec:models}.
Obviously the above parameters can only attain discrete values, and are characteristic  of a given messenger content. For these reasons, the Barbieri-Giudice fine tuning measure $\Delta_{\rm BG}$ \cite{Barbieri:1987fn} (see the next section for details) should not be computed with respect to them, once the messenger sector is specified.

In order to isolate the effect of non-unified messengers, we focus on a single source of SUSY breaking. In other words, we work with the ordinary parameters of minimal GM, i.e. the mediation scale and the SUSY-breaking F-term:
\begin{equation*}
M,~\Lambda\equiv \frac{F}{M},
\end{equation*}
as well as $\tan\beta$ and ${\rm sgn}(\mu)$.
In terms of the above parameters, the expressions for gaugino and scalar masses at the messenger scale read \cite{Martin:1996zb}: 
\begin{align}
\label{eq:soft1}
M_a (M) &= \frac{\alpha_a(M)}{4\pi} b_a^M \Lambda,~~a=1,2,3, \\
\tilde{m}^2_X (M) &= 2\sum_{a=1,3} \left(\frac{\alpha_a(M)}{4\pi}\right)^2 C^X_a b_a^M \Lambda^2,
\label{eq:soft2}
\end{align}
where $X$ refers to the MSSM superfields $X=Q,\,U,\,D,\,L,\,E,\,H_u,\,H_d$
and $C^X_a$ ($a=1,2,3$) is the quadratic Casimir of the representation of $X$ under $SU(3)\times SU(2) \times U(1)$.
We suppressed flavor indices of the sfermion mass matrix, as these contributions are flavor universal.
Unlike sfermion masses, A-terms are not generated at the leading order and are therefore highly suppressed
at the messenger scale.

To summarise, we are going to make a scan over GM models with non-unified messenger sectors, 
employing just two parameters more than in minimal GM: $b_1^M,~b_2^M,~b_3^M$, instead of the number of $\mathbf{5+\bar{5}}$ representations $N$.

\section{EWSB and fine tuning measures}
\label{sec:FT}
In our numerical analysis, we are going to adopt the following definitions of fine tuning, $\Delta_{\rm EW}$ and
$\Delta_{\rm HS}$, introduced in Refs.~\cite{Baer:2012up,Baer:2012mv}. They are defined starting from the minimization condition of the scalar potential
\begin{equation}
\label{eq:ewsb}
\frac{m_Z^2}{2} = \frac{(\tilde{m}^2_{H_d} +\Sigma_d) - (\tilde{m}^2_{H_u} +\Sigma_u)\tan^2\beta}{\tan^2\beta-1} - \mu^2,
\end{equation}
where all the parameters are evaluated at the electro-weak symmetry breaking (EWSB) scale. 
The quantitates $\Sigma_{u,d}$ arise from 1-loop corrections to the tree-level potential and their explicit forms can be
found in the Appendix of Ref.~\cite{Baer:2012cf}. 
The electro-weak fine tuning $\Delta_{\rm EW}$ is defined in terms of the 
sensitivity coefficients $C_x$ to the quantities appearing in the right-hand side of Eq.~(\ref{eq:ewsb})
(e.g.~$C_{H_d} \equiv \tilde{m}^2_{H_d} /(\tan^2\beta-1)$, 
$C_{\Sigma_d} \equiv \Sigma_d /(\tan^2\beta-1)$, 
$C_{H_u} \equiv - \tilde{m}^2_{H_u} \tan^2\beta/(\tan^2\beta-1)$, \dots) as follows \cite{Baer:2012up,Baer:2012mv}
\begin{equation}
\Delta_{\rm EW} \equiv \frac{\max_x |C_x|}{m^2_Z/2}. 
\end{equation}
The high-scale measure $\Delta_{\rm HS}$ is computed taking into account the RG running of the parameters to some input scale (in our case the messenger mass scale $M$),
$\tilde{m}^2_{H_{u,d}} = \tilde{m}^2_{H_{u,d}}(M) + \delta \tilde{m}^2_{H_{u,d}}$, $\mu^2 = \mu^2(M) + \delta \mu^2$ \cite{Baer:2012up,Baer:2012mv}
\begin{equation}
\Delta_{\rm HS} \equiv \frac{\max_x |B_x|}{m^2_Z/2},
\end{equation}
where $B_{H_d} \equiv \tilde{m}^2_{H_d}(M) /(\tan^2\beta-1)$, 
$B_{\delta_{H_d}} \equiv \delta \tilde{m}^2_{H_d} /(\tan^2\beta-1)$, etc.

Later, we will compare the resulting values of $\Delta_{\rm EW}$ and $\Delta_{\rm HS}$ for some representative points of the parameter space with the classical definition by Barbieri and Giudice, $\Delta_{\rm BG}$ \cite{Barbieri:1987fn}:
\begin{equation}
\Delta_{\rm BG} \equiv \max_x \left|\frac{\partial \log m_Z^2}{\partial \log a_x}\right|,
\end{equation}
where $a_x$ are the fundamental high-energy parameters, 
in our case: $\Lambda$, $M$, $\tan\beta$, $\mu^2$.

We will show that, given the high degree of correlation of the soft masses 
(the spectrum is controlled by the single dimensionful high-energy parameter $\Lambda = F/M$), 
$\Delta_{\rm HS}$ seems to badly overestimate the fine tuning of our non-unified models, 
while $\Delta_{\rm EW}$ and  $\Delta_{\rm BG}$ return values of the similar order, 
much lower than $\Delta_{\rm HS}$~\cite{Baer:2013gva}. 
The reason of such an overestimate is that $\Delta_{\rm HS}$ does not take into account automatic cancellations 
that arise from the fact that terms in $B_x$ are related to a few fundamental parameters.\footnote{See
Ref.~\cite{Mustafayev:2014lqa} for detailed discussion of $\Delta_{\rm EW}$, $\Delta_{\rm HS}$ and $\Delta_{\rm BG}$.}

We will therefore adopt $\Delta_{\rm EW}$ as our main indication of naturalness and, in particular, 
we will show how models with $b^M_2 > b^M_3$ can be considerably less tuned than minimal GM, {\it i.e.}, 
in set-ups with unified messengers, $b^M_1= b^M_2 = b^M_3$.

The above sketched outcome generalizes to the case of gauge mediation the results of 
\cite{Horton:2009ed,Antusch:2012gv,Kaminska:2013mya}. 
It can be intuitively understood considering the minimization condition in Eq.~(\ref{eq:ewsb}). 
As is well known, for moderate to large values of $\tan\beta$, the equation reduces to
\begin{equation}
\label{eq:ewsb2}
\frac{m_Z^2}{2} \approx - (\tilde{m}^2_{H_u} +\Sigma_u) - \mu^2,
\end{equation}
such that the EWSB can be achieved if $\tilde{m}^2_{H_u}<0$ and the correct $Z$-boson mass is obtained by cancellation
mainly between by $|\tilde{m}_{H_u}|$ and $\mu$, unless the radiative corrections encoded in $\Sigma_u$ dominate (as in the case of very heavy sfermions). The larger the values of these parameters, the more severe the tuning will result. The low-energy value of $\tilde{m}^2_{H_u}$ is related to that generated at the messenger scale by the renormalization group equation (RGE) whose expression at one loop is:
\begin{equation}
\label{eq:rge}
16\pi^2 \frac{d}{dt} \tilde{m}^2_{H_u} \simeq 6 y_t^2 (\tilde{m}^2_{H_u} + (\tilde{m}^2_{Q})_{33}+
 (\tilde{m}^2_{U})_{33}) + 6 A_t^2 - 6g_2^2M_2^2 - \frac{6}{5}g_1^2M_1^2- \frac{3}{5}g_1^2 S\,,
\end{equation}
where $t$ is related to the renormalization scale as $t\equiv \log(\mu/M)$, $y_t$ and $A_t$ are 
top Yukawa and A-term, and 
$S\equiv \tilde{m}^2_{H_u} - \tilde{m}^2_{H_d} + {\rm Tr}( \tilde{m}^2_{Q}-\tilde{m}^2_{L}-2 \tilde{m}^2_{U}
+\tilde{m}^2_{D}+\tilde{m}^2_{E})$.

As is well known, negative values of $\tilde{m}^2_{H_u}$ (and thus the EWSB) can be radiatively induced by the contributions $\propto y_t^2$ of the stop mass parameters $(\tilde{m}^2_{Q})_{33}$ and $(\tilde{m}^2_{U})_{33}$, which tend to decrease the value of $\tilde{m}^2_{H_u}$ in the running from $M$ to low energies. The large stop masses, as required 
by the observed Higgs mass in minimal GM, make $\tilde{m}^2_{H_u}$ run to more negative values, thus giving at
low energy $|\tilde{m}^2_{H_u}| \gg m_Z^2$, which results in a severe fine tuning. This is also the effect of heavy gluinos,
given that $(\tilde{m}^2_{Q})_{33}$ and $(\tilde{m}^2_{U})_{33}$ grows with $M_3$ in the running.
On the other hand, the $SU(2)$ and $U(1)$ gaugino contributions $\propto g_i^2 M_i^2$ 
in the RGE in Eq.~(\ref{eq:rge}) tend to compensate the above described effects, making $\tilde{m}^2_{H_u}$ grow in the
running to low energies. While this effect is subdominant for universal gaugino masses, it can substantially reduce
the low-energy value of $|\tilde{m}^2_{H_u}|$ even for heavy stops and gluinos, if $M^2_{1,2}> M^2_{3}$ at high energies. Because of the hierarchy of the gauge couplings, as well as the numerical coefficients in  Eq.~(\ref{eq:rge}),
this compensation is much more effective in the Wino case than for the Bino.

The above considerations make us expect a reduced tuning for those models where the gauge-mediated SUSY breaking gives $M^2_{2}\gg M^2_{3}$ at the mediation scale $M$, i.e.~whose messenger sector corresponds to $b^M_2 > b^M_3$.
In the following, we are going to demonstrate this effect by means of a numerical scan over these kinds of models.   

\section{Numerical scan and fine tuning}
\label{sec:numerics}
We performed a scan of the parameter space described above in section \ref{sec:model}, employing a version of 
{\tt ISAJET 7.85} \cite{Baer:1999sp}, to which we added a subroutine to compute $\Delta_{\rm BG}$.
For the messenger contributions to the $\beta$-function coefficients, we took random integer values~\footnote{For
$b_2^M$ and $b_3^M$, this choice is justified by the fact that a fundamental representation of $SU(N)$
  has index 1/2, and messengers come in vector-like representations of the SM gauge group. $b_1^M$ can instead
  attain fractional values as $n/5$ with $n$ an integer. See section \ref{sec:models} for further discussion.}  
 within these intervals
\begin{align}
1 \le &~ 5\times b_1^M\le 75, \nonumber\\
1 \le &~ b_2^M \le 15, \nonumber\\
1 \le &~ b_3^M \le 7.
\label{eq:branges}
\end{align}
The other parameters are randomly varied in the following ranges
\begin{align}
5\times10^4~{\rm GeV} \le  ~\Lambda \le 10^6 ~{\rm GeV}, \nonumber & \\
2\times \Lambda \le  ~M   \le 10^{15} ~{\rm GeV}, \nonumber &\\
5 \le  ~\tan\beta ~ \le 50, & \label{eq:ranges}
\end{align}
and the top mass was set to $m_t = 173.3$ GeV \cite{ATLAS:2014wva}.
We checked that ${\rm sgn}(\mu)$ has a very mild impact on the fine-tuning measures when the other
parameters are fixed, hence we employed $\mu>0$ throughout the paper.

For the sake of comparison with minimal GM, we also performed a scan within the same ranges of the parameters, but with unified messengers: 
\begin{equation}
\label{eq:mGM}
b^M_1= b^M_2 = b^M_3 \equiv b^M,\quad 1 \le  b^M \le 7.
\end{equation}
\begin{figure}[t]
\begin{center}
\includegraphics[width=0.49\textwidth]{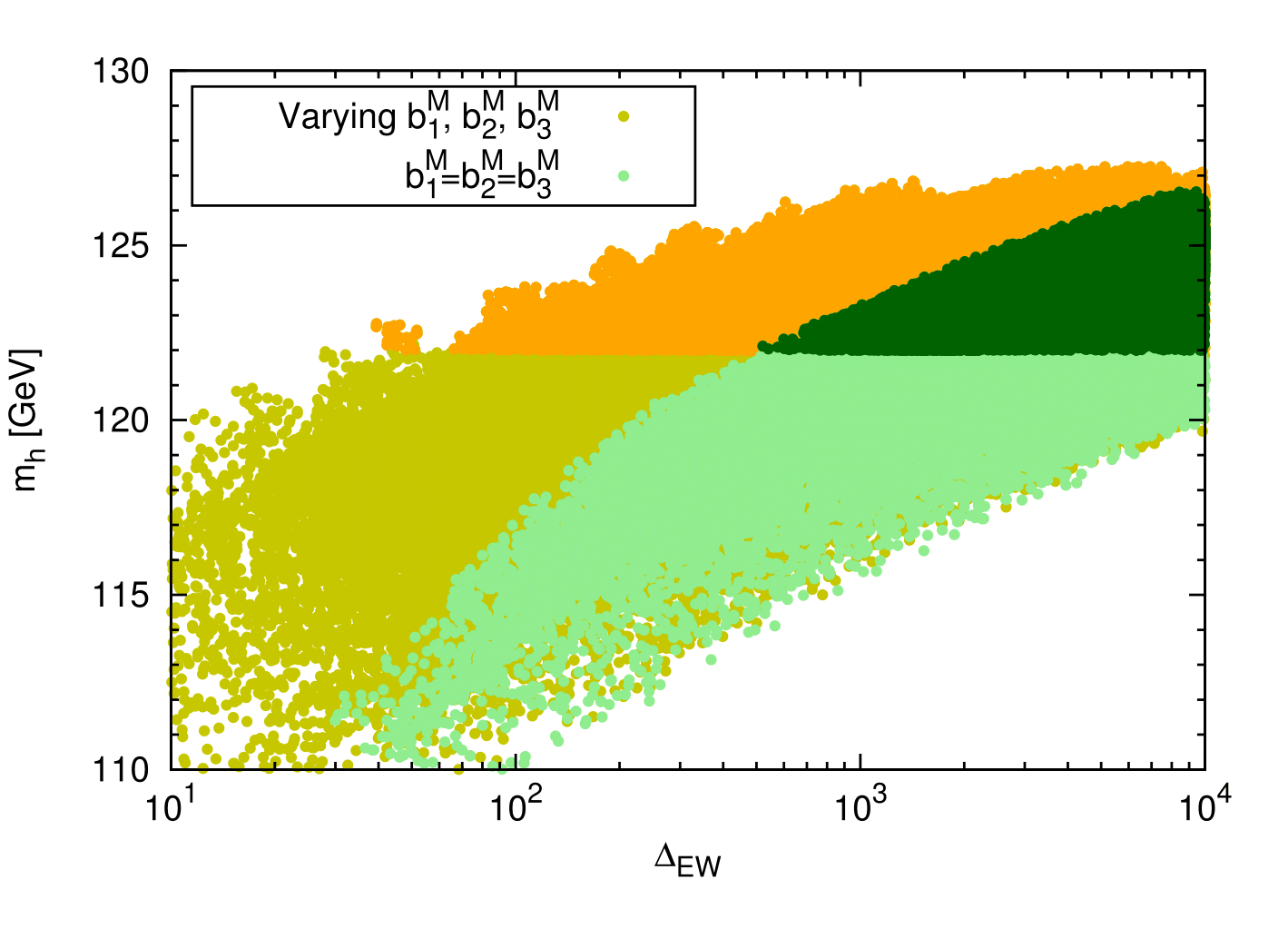}
\includegraphics[width=0.49\textwidth]{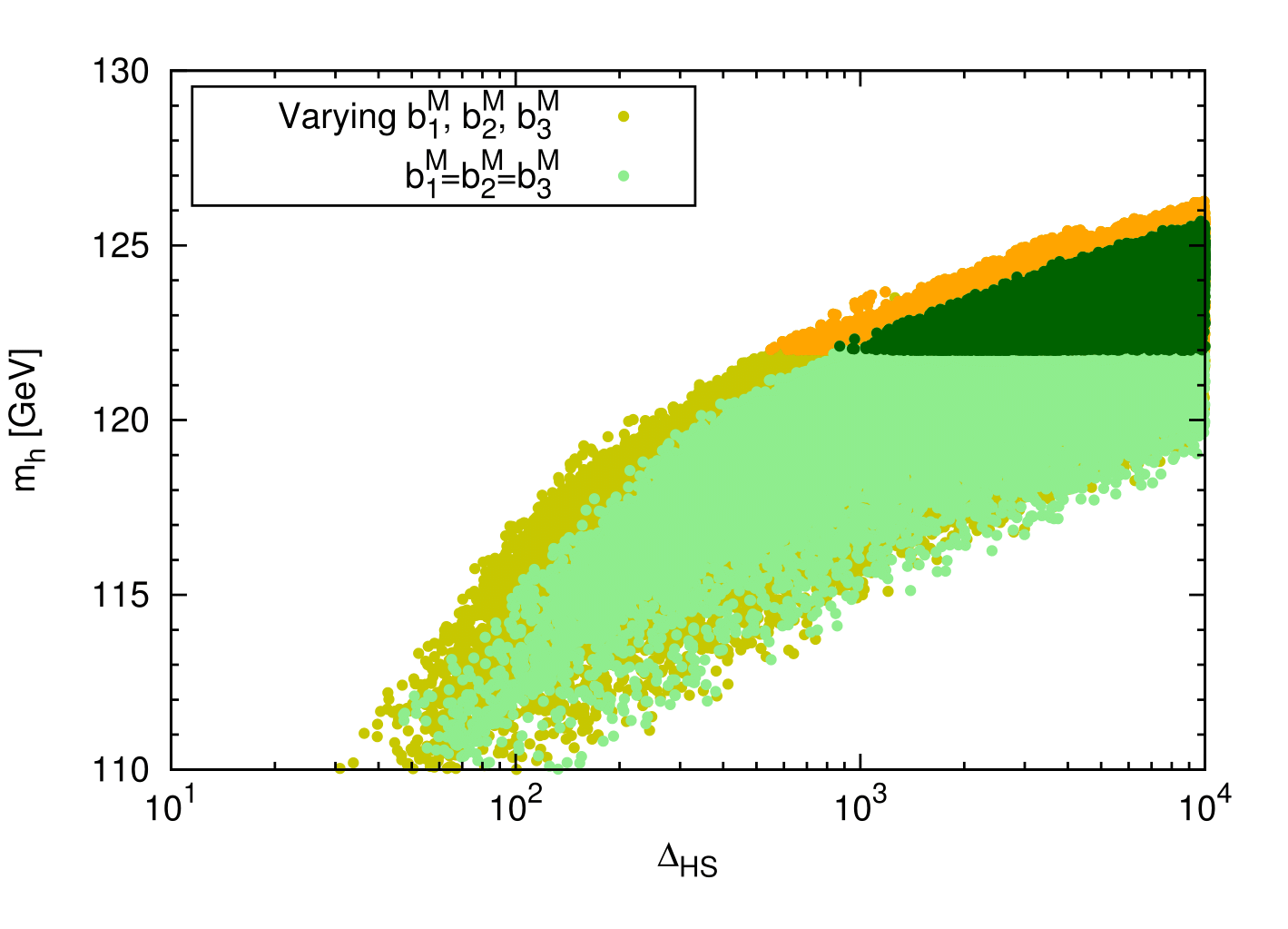}
\caption{Higgs mass vs.~the electroweak fine tuning $\Delta_{\rm EW}$  (left) and the high-scale measure 
$\Delta_{\rm HS}$  (right) for the scan of Eq.~(\ref{eq:branges}) over models with non-unified messengers (dark-yellow and orange points), compared to the scan of the minimal GM parameter space as in Eq.~(\ref{eq:mGM}) (green and dark-green points). See the text for details.
\label{fig:mh-FT}}
\end{center}
\end{figure}
Our results are displayed in Fig.~\ref{fig:mh-FT}. The left panel shows the light Higgs mass $m_h$ and the fine tuning $\Delta_{\rm EW}$ for those points that survive the basic consistency checked by {\tt ISAJET} (no tachyons, correct EWSB, etc.). Yellow points correspond to the scan over the models with non-unified messengers as defined 
in Eq.~(\ref{eq:branges}), green points to the minimal GM scan of Eq.~(\ref{eq:mGM}). The orange and dark-green points
highlight the solutions with $m_h$ compatible with the measured value within experimental and theoretical uncertainties (which we take to amount to 3 GeV):
\begin{equation}
\label{eq:mh}
122~{\rm GeV} \le  m_h \le 128~{\rm GeV}.
\end{equation}
\begin{figure}[t]
\begin{center}
\includegraphics[width=0.49\textwidth]{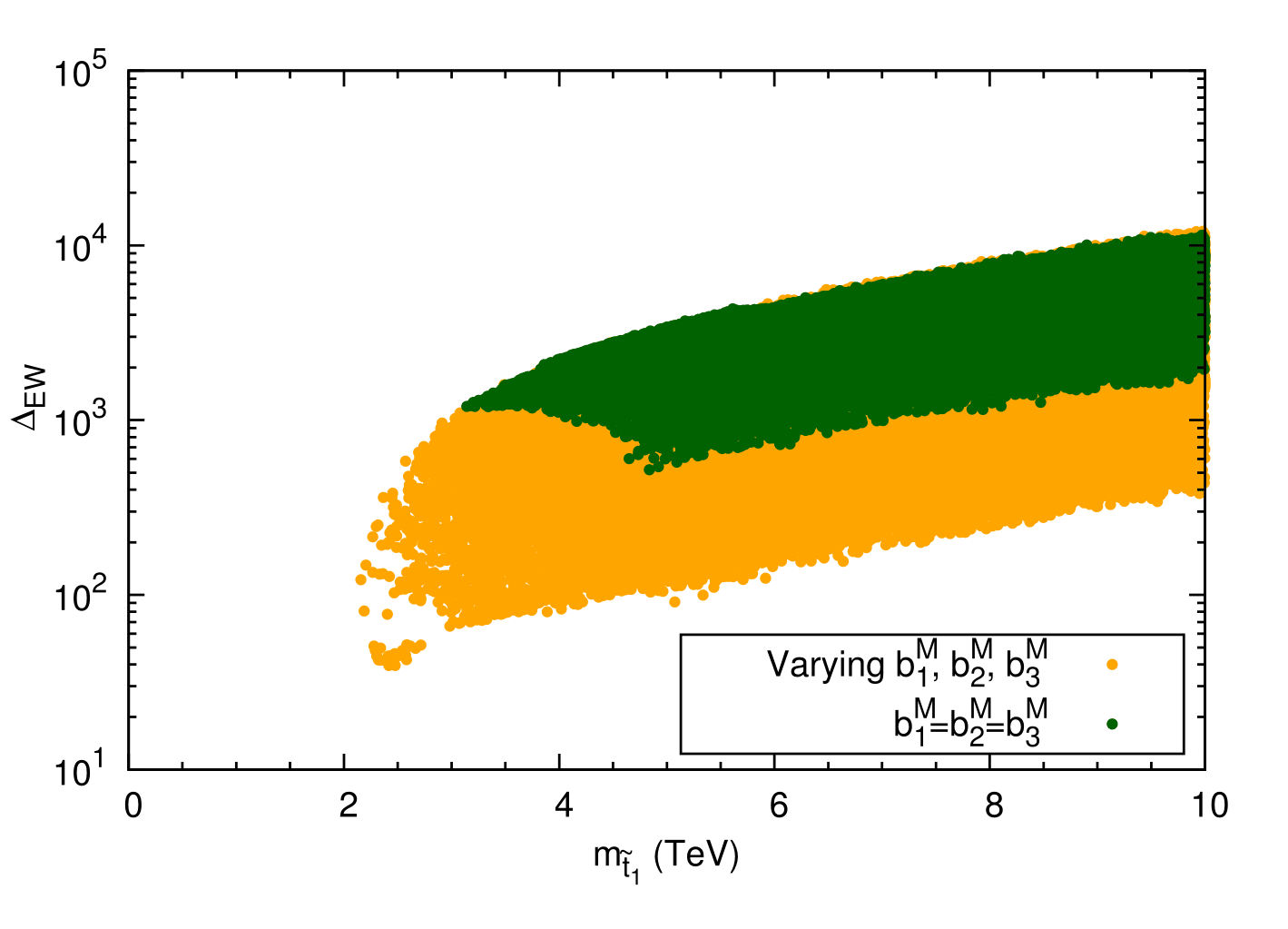}
\includegraphics[width=0.49\textwidth]{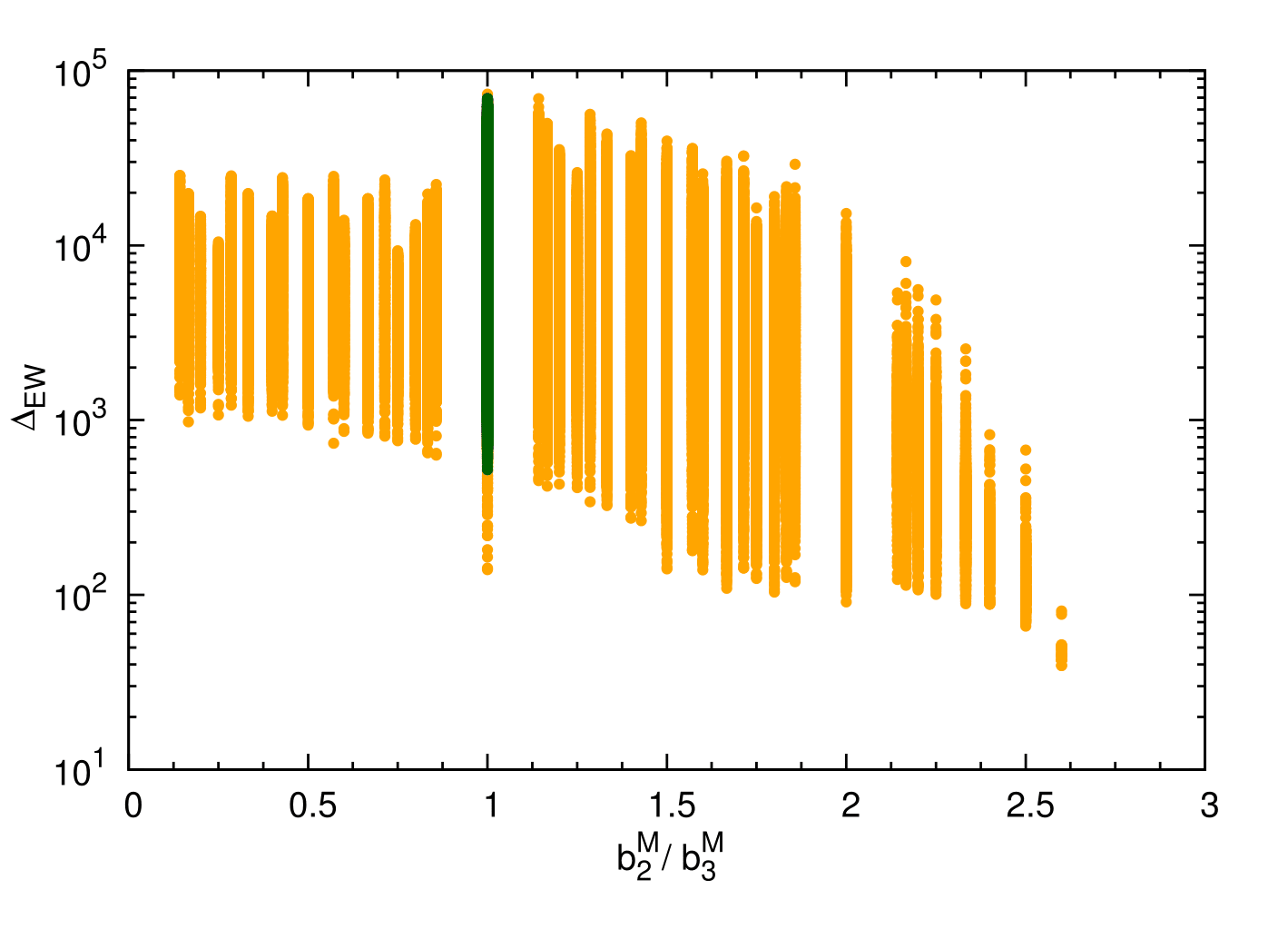}
\caption{$\Delta_{\rm EW}$ vs.~the lightest stop mass (left) and the ratio of $\beta$-function coefficients $b_2^M/b_3^M$ (right)
for points satisfying $122~{\rm GeV} \le  m_h \le 128~{\rm GeV}$. As in the previous figure, orange points correspond to 
models with non-unified messengers, while dark-green points to minimal GM, i.e.~$b^M_1= b^M_2 = b^M_3$. 
\label{fig:comp} }
\end{center}
\end{figure}
In Fig.~\ref{fig:comp}, we show $\Delta_{\rm EW}$  vs.~the lightest stop mass (left), as well as a function of the ratio
$b_2^M/b_3^M$, only for the points corresponding to the above Higgs mass range.
From these plots, we can see that the condition of Eq.~(\ref{eq:mh}) 
constrains $\Delta_{\rm EW} \gtrsim 500$ for minimal GM. This well-known result reflects the heavy stop masses required to raise the Higgs mass at the level of Eq.~(\ref{eq:mh}), given the vanishing values of the stop A-term
generated at the messenger scale in minimal GM, see e.g.~\cite{Gogoladze:2015tfa,Ajaib:2012vc}.\footnote{However, within the context of General Gauge Mediation \cite{Meade:2008wd}, there are ways to generate a sizeable $|A_t|$ through the RG running, so that $m_{{\tilde t}_1}\lesssim 1$ TeV can be still compatible with $m_h=125$ GeV \cite{Knapen:2015qba}.} 
On the other hand, in the case of non-unified messengers the $\Delta_{\rm EW}$ can be as low as $40\div50$, corresponding to a substantial reduction of the fine tuning (about one order of magnitude). 
This confirms our qualitative considerations at the end of section \ref{sec:FT}: it is a consequence of 
the compensating effects of the Wino mass $M_2$ and the stop masses in the running of $m^2_{H_u}$.
In fact, lowest values of $\Delta_{\rm EW}$ are achieved for the largest possible values of the ratio $b_2^M/b_3^M$
giving acceptable solutions, cf.~Fig.~\ref{fig:comp}, right. The larger values of $b_2^M/b_3^M$ give 
${\tilde m}^2_{H_u} >0$ at low energies, hence failing to trigger a successful EWSB, because of the $M_2^2$ term dominating the RGE in~Eq.~(\ref{eq:rge}).

In the right panel of Fig.~\ref{fig:mh-FT}, we display the result of our scans in terms of the high-energy measure 
$\Delta_{\rm HS}$. As we can see $\Delta_{\rm HS}$ does not benefit from a substantial improvement with respect to minimal GM in models with non-unified messengers. However, $\Delta_{\rm HS}$ does not take into account correlations of the parameters in the fundamental high-energy theory, hence likely overestimates fine tuning, as
argued in  \cite{Baer:2013gva}. We are going to support this conclusion in the next section by comparing $\Delta_{\rm EW}$ and $\Delta_{\rm HS}$ with the Barbieri-Giudice measure $\Delta_{\rm BG}$.

\section{Typical spectra of low-tuned models}
\label{sec:spectrum}
In this section, we discuss the phenomenological aspects of models of gauge mediation with non-unified messengers 
that can feature a low degree of tuning, starting from the typical spectrum. 
From the discussion of the previous sections, we can already anticipate some features of this kind of models. 
First of all, we expect a heavy stop sector and a heavy gluino to raise the Higgs mass at the observed value, 
as in any GM model without extra contributions to generate large A-terms at the mediation scale. 
Second, a low value of the Higgsino mass $\mu$ is obviously needed if we require 
no fine cancellations in Eq.~(\ref{eq:ewsb}). On the other hand, large Wino and possibly Bino masses 
are the key ingredient of our non-unified models to keep $|m^2_{H_u}|$ small at the EWSB scale, despite the heavy stops, thanks to the compensating effect in the RGE of Eq.~(\ref{eq:rge}). Hence, everything that is charged under SU(2) and SU(3) will be definitely heavy, while the lightest chargino  and the two lightest neutralinos are likely to be Higgsino-like with possibly a sizeable Bino component for solutions with $M_1 \simeq \mu$.  
To summarize, besides the usual gravitino LSP, we expect that the only light states will be the Higgsinos (thus likely to provide a neutralino NLSP) and possibly Binos and RH sleptons (especially the RH stau). 
\begin{figure}[t]
\begin{center}
\includegraphics[width=0.49\textwidth]{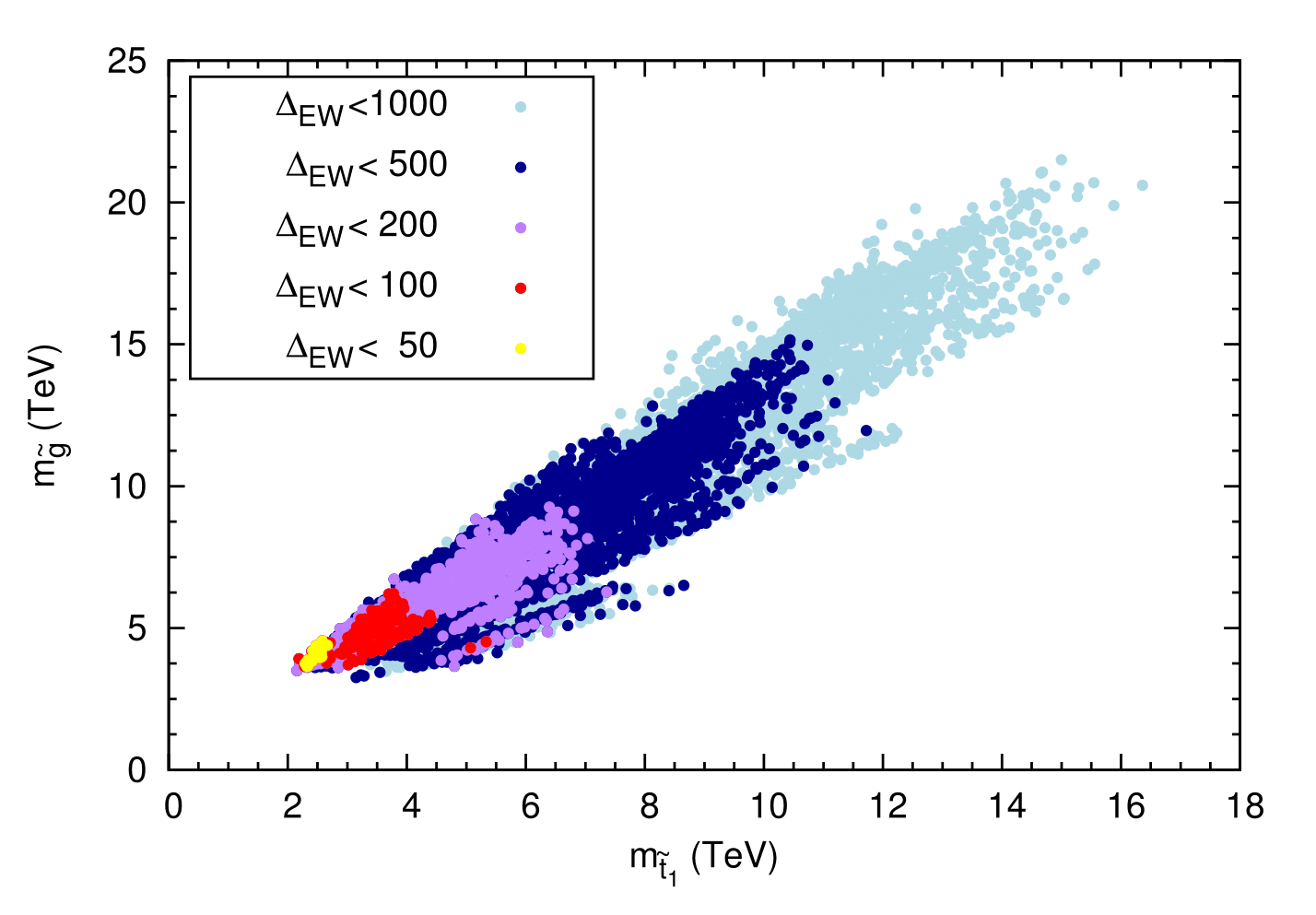}
\includegraphics[width=0.49\textwidth]{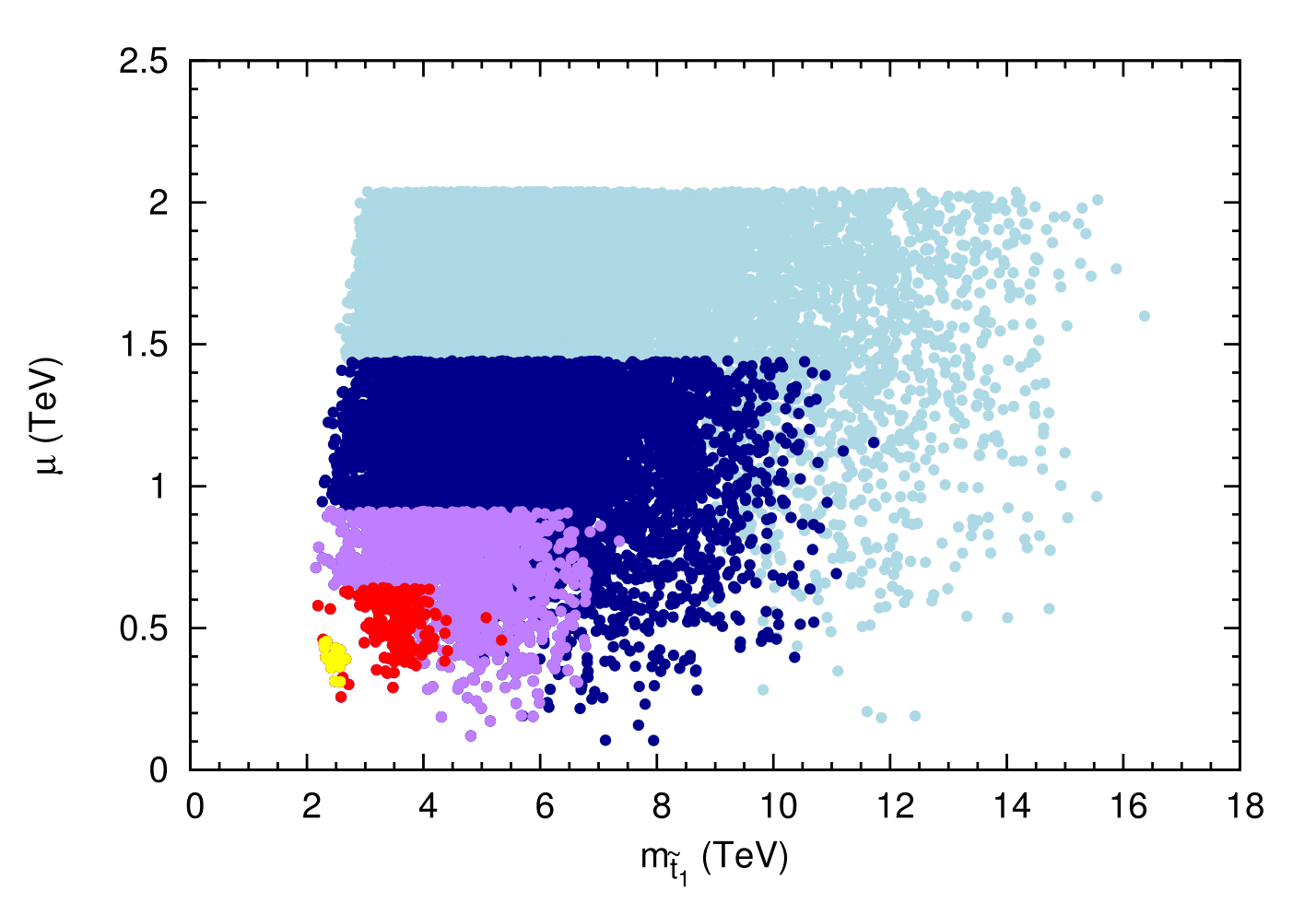}
\caption{Gluino mass (left) and Higgsino mass (right) vs.~the lightest stop mass for the models 
of non-unified messengers corresponding to the lowest values of $\Delta_{\rm EW}$. All points satisfy Eq.~(\ref{eq:mh}).
\label{fig:spectrum1} }
\end{center}
\end{figure}

In Fig.~\ref{fig:spectrum1}, we show the points of our scan defined in Eqs.~(\ref{eq:branges}, \ref{eq:ranges})
that fulfil the Higgs mass range of Eq.~(\ref{eq:mh}) and feature $\Delta_{\rm EW} < 1000$. The lower fine-tuning ranges are displayed in different colors as indicated in the figure. In the left panel, we show the result in the plane of the physical lightest stop and gluino masses, $(m_{\tilde{t}_1},\,m_{\tilde{g}})$. In the right panel, we show the lightest stop mass and the Higgsino mass parameter $\mu$, $(m_{\tilde{t}_1},\,\mu)$. 
As expected the color superpartners are rather heavy: solutions are only found for 
$m_{\tilde{t}_1}\gtrsim 2$~TeV, $m_{\tilde g}\gtrsim 3$~TeV. Remarkably, very heavy squarks and gluinos, $\mathcal{O}(10)$~TeV,  are still compatible with a fine-tuning better than the permil level. However, the points with lowest tuning,     
$\Delta_{\rm EW} < 50$, requires $m_{\tilde{t}_1}\lesssim 2.5$~TeV, $m_{\tilde g}\lesssim 5$~TeV, to avoid too large radiative correction $\Sigma_u$ in Eq.~(\ref{eq:ewsb2}). 
On the other hand, Higgsinos need to be lighter than about 500 GeV for $\Delta_{\rm EW} < 50$.
\begin{figure}[t]
\begin{center}
\includegraphics[width=0.49\textwidth]{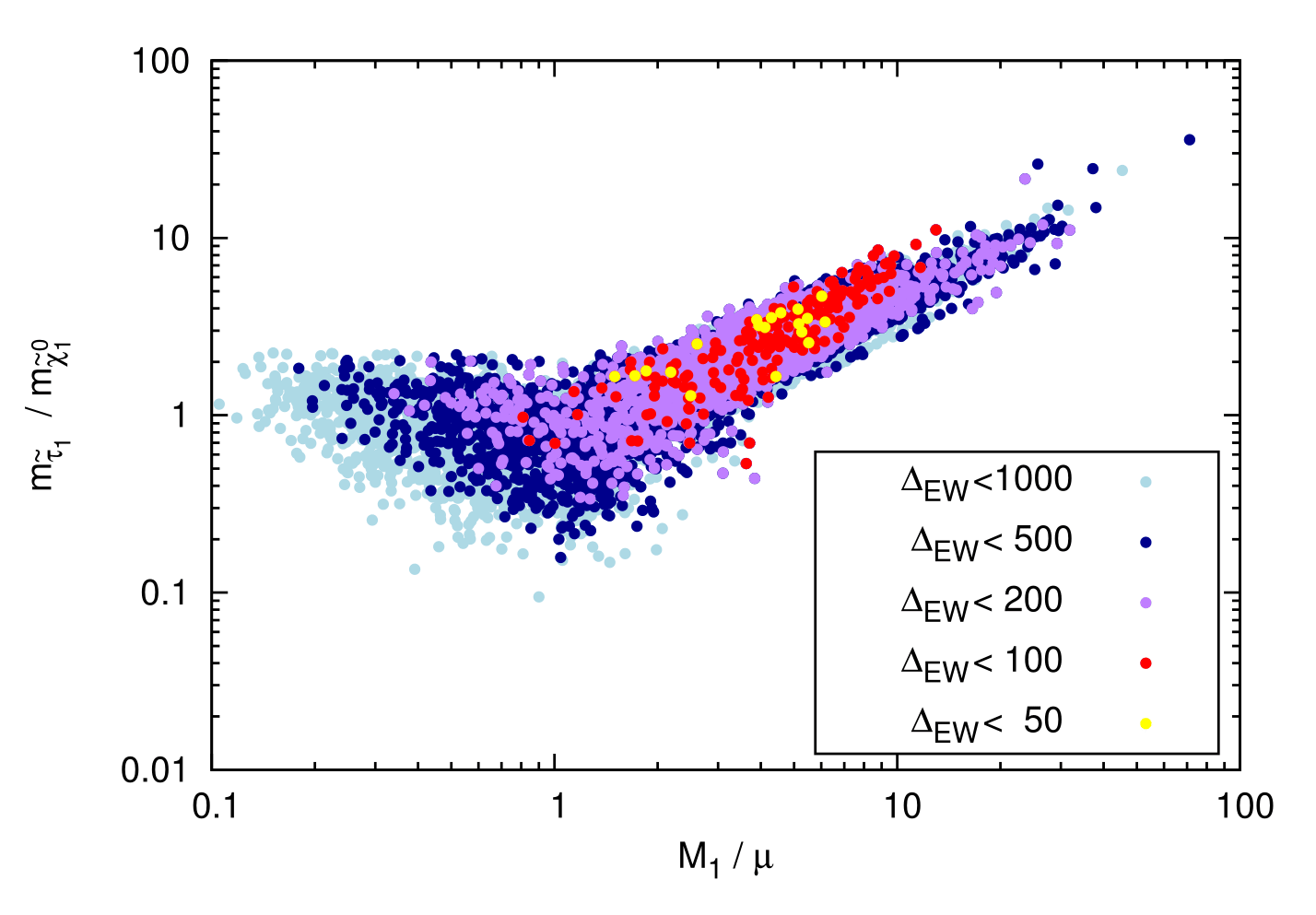}
\includegraphics[width=0.49\textwidth]{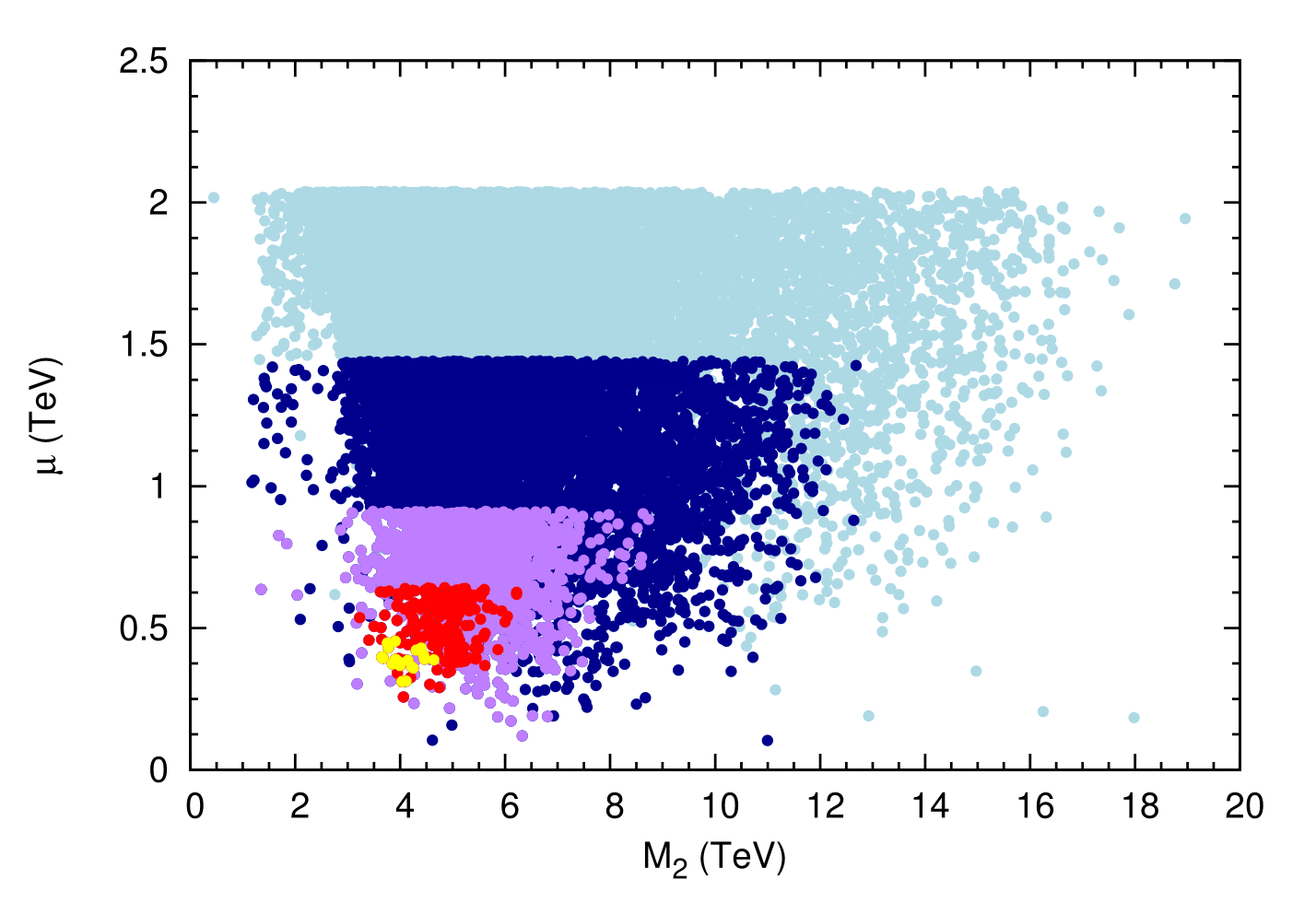}
\caption{The same as in the previous figure, for the parameter 
ratios $m_{\tilde{\tau}_1}/m_{\tilde{\chi}^0_1}$ vs.~$M_1/\mu$ (left) and
the Higgsino mass vs.~the Wino mass (right).
\label{fig:spectrum2} }
\end{center}
\end{figure}

Further features of low-tuned models are better depicted in Fig.~\ref{fig:spectrum2}. Scan and color code are the same as in the previous figure. The left panel of Fig.~\ref{fig:spectrum2} allows us to investigate nature and composition of the NLSP: on the $y$-axis we display the mass ratio of the lightest stau and neutralino, $m_{\tilde{\tau}_1}/m_{\tilde{\chi}^0_1}$, on the $x$-axis the ratio of the key parameters of the neutralino mass matrix, $M_1/\mu$. As we can see, solutions with
the lowest tuning always feature a neutralino NLSP, $m_{\tilde{\tau}_1}>m_{\tilde{\chi}^0_1}$, although a corner of the
parameter space with relatively low tuning, $\Delta_{\rm EW} < 100\div200$, features a (RH) stau NLSP. The lightest neutralino NLSP is always Higgsino-like ($M_1>\mu$) for  $\Delta_{\rm EW} < 50$, but we also see some points with
$M_1\simeq\mu$ that translate to a sizeable Bino-Higgsino mixing. 

The right panel of Fig.~\ref{fig:spectrum2} shows the Wino-Higgsino mass plane. As we expected, Winos have to be heavy, $M_2 \gtrsim 2$~TeV, and low-tuning even requires $M_2 \approx 4$~TeV. Thus, the lightest chargino will be almost pure Higgsino and negligible Wino component is expected in the lightest neutralinos, as well. 
\begin{figure}[t]
\begin{center}
\includegraphics[width=0.49\textwidth]{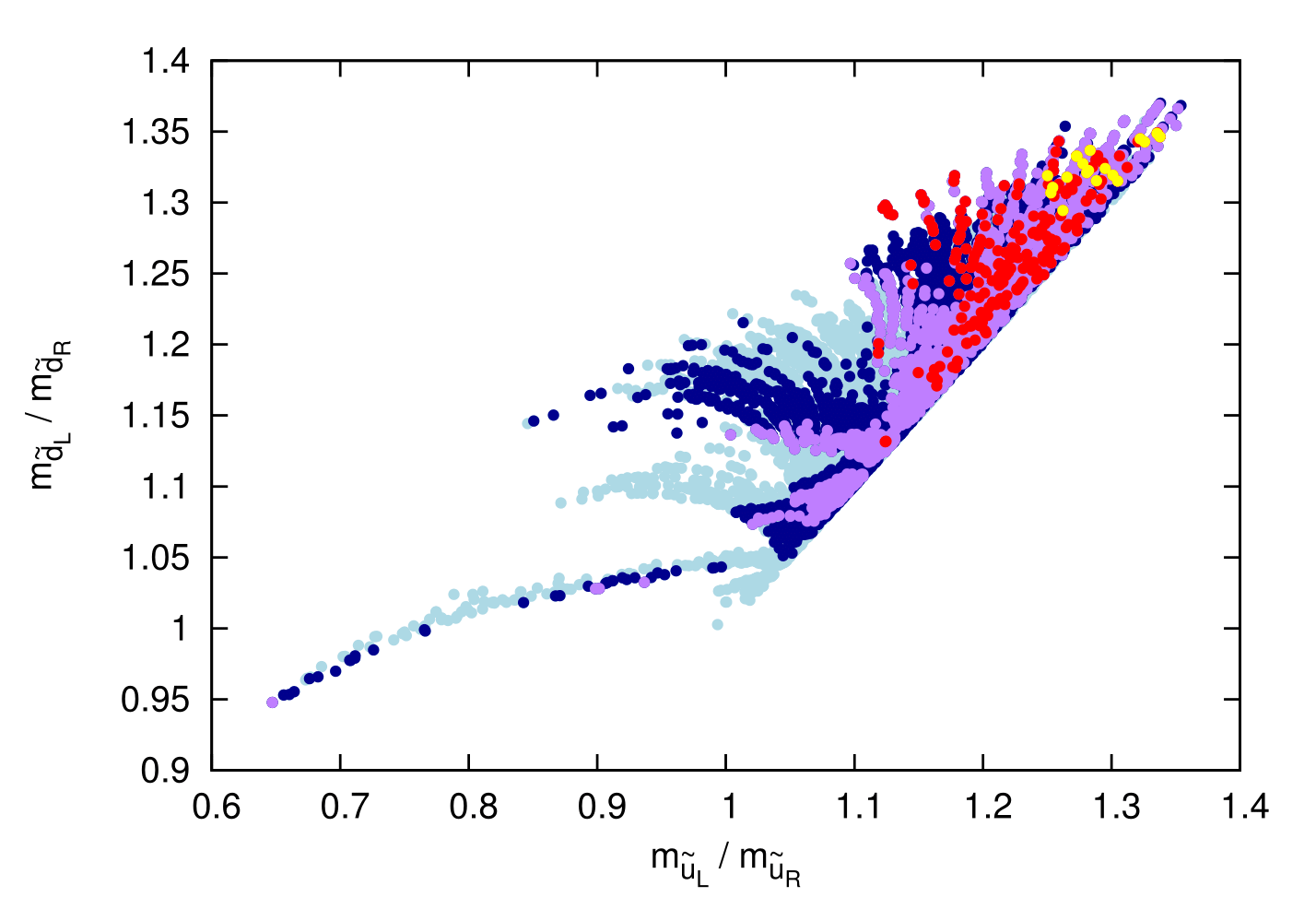}
\includegraphics[width=0.49\textwidth]{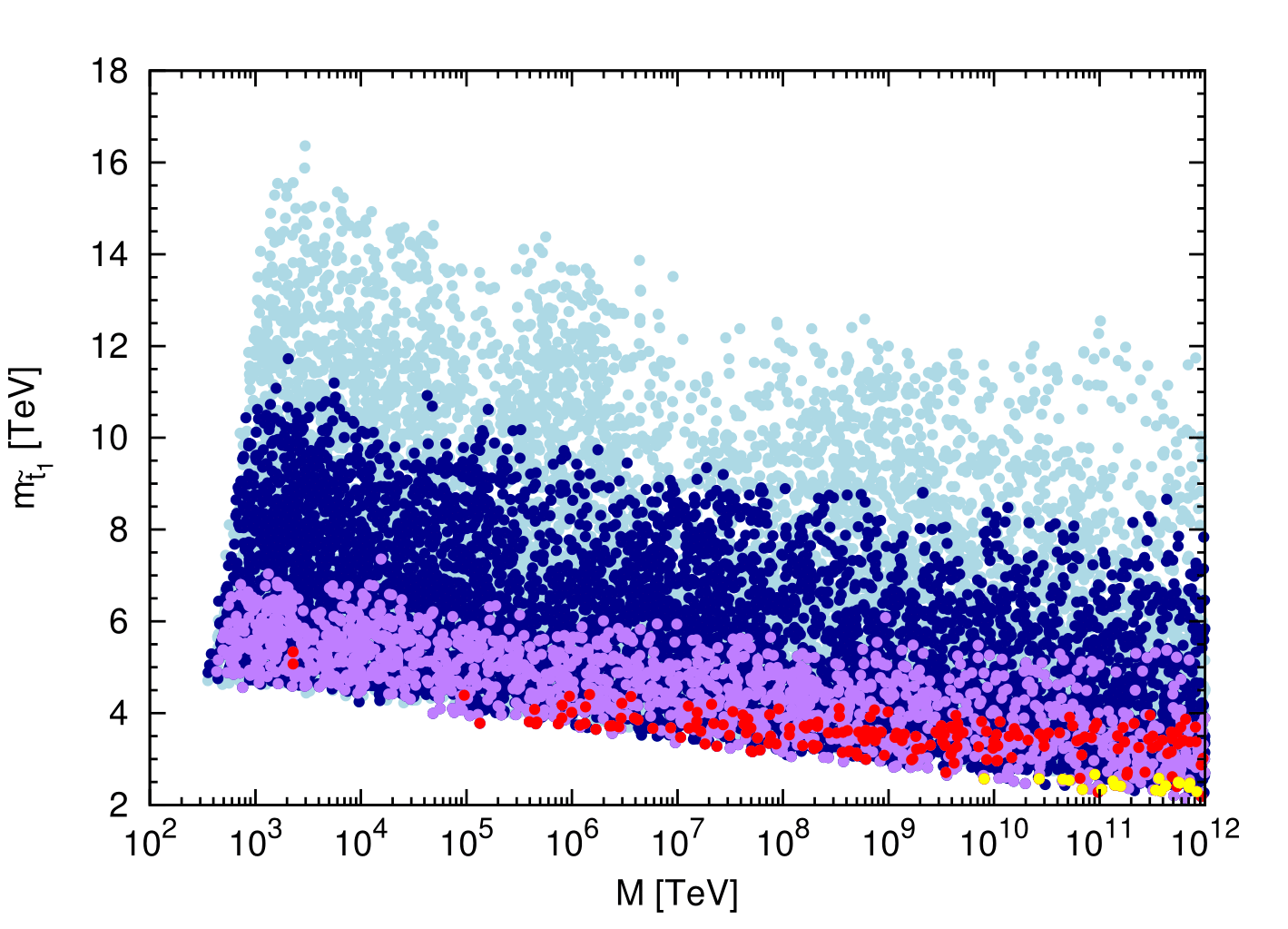}
\caption{
The same as in the previous figures, for the squark mass ratios $m_{\tilde{d}_L}/m_{\tilde{d}_R}$ vs.~$m_{\tilde{u}_L}/m_{\tilde{u}_R}$ (left) and the lightest stop mass vs.~the mediation scale $M$ (right).
\label{fig:spectrum3} }
\end{center}
\end{figure}

The left panel of Fig.~\ref{fig:spectrum3} displays the first generation squark mass ratios, ($m_{\tilde{u}_L}/m_{\tilde{u}_R}$, $m_{\tilde{d}_L}/m_{\tilde{d}_R}$). These particles are in the multi-TeV range, as we have seen before. However, their ratios show the imprint of the non-universal gaugino masses. In fact, the low-tuning solutions feature left-handed (LH) squarks
(i.e.~$SU(2)$ doublets) about 30\% heavier than the right-handed (RH) ones ($SU(2)$ singlets), as a consequence of the large Wino contribution $\propto g_2^2 M_2^2$ in the squark mass running. However, few points with $\Delta_{\rm EW} < 200$
feature the opposite hierarchy: this reflects some relatively low-tuned solutions with $M_1\gg M_2,~M_3$.
The structures we observe in the plot arise from the fact that we scan over discrete values of $b_1^M$.

Finally, in the right panel of Fig.~\ref{fig:spectrum3}, we show the lightest stop mass vs.~the mediation scale $M$. 
First of all, we see that the stop mass can be lower for a high mediation scale. This reflects the well-known fact that 
a longer running generates a sizeable $A_t$ for heavy enough gluinos, so that a lighter stop sector can provide the
required radiative correction to the Higgs mass due to the stop LR mixing effects, see e.g.~\cite{Grajek:2013ola}. 
Secondly, we see that the fine-tuning
prefers large $M$ as well. This is not only because stops can be lighter -- which diminishes the radiative correction $\Sigma_u$ in Eq.~(\ref{eq:ewsb}) -- but also because a longer running makes the compensation of stop and Wino contributions in Eq.~(\ref{eq:rge}) more efficient. 
As we are going to discuss in section \ref{sec:collider}, a high mediation scale translates to a long-lived NLSP
compared to the detector scale at collider experiments.
Only few points with $\Delta_{\rm EW} < 100$ correspond to a low mediation scale $10^{3\div4}$~TeV, thus being
characterized by an NLSP decaying inside the detector.
\subsection{Benchmark models}
\begin{table}[t]
{\small
\begin{center}
\begin{tabular}{@{} |c||c|c|c|c|c| @{}}
\hline    \hline    
     & A & B & C & D & E \\ 
\hline    
    $\Lambda$ & 135 TeV & 215 TeV & 120 TeV  & 211 TeV &  314 TeV \\ 
    $M$   & $2 \times10^{13}$ & $5 \times10^{11}$ & $9.2\times10^{14}$ &  $3.7\times10^{11}$  & $2.3\times10^{6}$ \\ 
    $\tan\beta$ & 20 & 15  & 14 & 25  & 18 \\ 
    $(b^M_1,~b^M_2,~b^M_3)$ & ($\frac{19}{5}$, 13, 5) & ($\frac{8}{5}$, 10, 4) & ($\frac{19}{5}$, 13, 5)  & ($\frac{14}{5}$, 10, 4)  & (2, 4, 2) \\ 
\hline    
    $\mathbf{\Delta_{\rm EW}}$ & {\bf 48} & {\bf 96}  & {\bf 81} & {\bf 84}  & {\bf 90} \\ 
    $\Delta_{\rm HS}$ & 771 & 1427 & 710 & 1365  & 1076 \\ 
    $\mathbf{\Delta_{\rm BG}}$ & {\bf 87} & {\bf 122} & {\bf 98} & {\bf 104}  & {\bf 95}  \\ 
    \hline    
        $\mu$   & 314 & 632 & 579 & 554  & 610 \\ 
        $m_h,~m_A$   & 122, 3207 & 122,3880 & 122, 3255 & 123, 3535  & 122, 2350 \\ 
        \hline    
        $m_{\tilde{g}}$   & 4342 & 5439 & 3901 & 5343  & 4178  \\ 
    $m_{\tilde{\chi}^0_{1,2}}$   & \textcolor{red}{320}, 325 &  \textcolor{red}{455}, 651 & \textcolor{red}{571}, 594  & 566, 572  & \textcolor{red}{621}, 628 \\ 
     $m_{\tilde{\chi}^0_{3,4}}$   & 700, 4427 & 654, 5405 & 639, 3935 & 809, 5307  & 874, 3233 \\ 
    $m_{\tilde{\chi}^\pm_{1,2}}$   & 335, 4405 & 669, 5376 &  610, 3915 & 589, 5278 & 644, 3202 \\     
    $m_{\tilde{t}_{1,2}}$  & 2644, 4683 & 3618, 5835 & 2178, 4322 &  3597, 5677 &  3941, 4766 \\     
    $m_{\tilde{b}_{1,2}}$  & 3777, 4701 & 4878, 5845 & 3456, 4347 & 4684, 5685  & 4393, 6190 \\     
    $m_{\tilde{\tau}_{1,2}}$  & 597, 3384 & 540, 3941 & 745, 3271 & \textcolor{red}{533}, 3834  & 794, 2338 \\     
    $m_{\tilde{u}_L},\,m_{\tilde{d}_L}$  & 5113, 5113 & 6277, 6278 & 4748,4749 & 6152,6153   &  4963, 4964  \\     
     $m_{\tilde{u}_R},\,m_{\tilde{d}_R}$  & 3920, 3892 & 4960, 4944 & 3554, 3520 & 4890, 4861  &  4449, 4424 \\     
    $m_{\tilde{\ell}_L},\,m_{\tilde{\ell}_R}$  & 3411, 790 & 3960, 603 & 3284, 819 & 3878, 887  & 2353, 799 \\     
    \hline
    $m_{\widetilde G}$  & 640 MeV & 25 MeV &  26 GeV & 19 MeV  & 175 eV \\ 
 \hline  \hline    
  \end{tabular}
 \end{center}
 }
  \caption{Spectrum and parameters of five benchmark models. Model A corresponds to characteristic low-tuned solution, featuring pure Higgsino NLSP. Models B and C give examples of Bino-like and mixed Higgino-Bino NLSP respectively.
 Models D and E show peculiar solutions with long-lived stau NLSP (model D) 
 and short-lived neutralino NLSP (model E).
 Dimensionful quantities are in GeV unless otherwise indicated. \label{tab:benchmarks}}
\end{table}
 
We close this section showing in Table \ref{tab:benchmarks} the full spectrum of five models with low fine tuning that
exemplify the features discussed above. 
The gravitino LSP mass is given in the last row, while the NLSP mass is highlighted in red. 

Model A corresponds to the typical low-tuned solution with a $\approx 300\div400$~GeV Higgsino and thus a Higgsino-like neutralino NLSP, almost degenerate to the second neutralino and the light chargino. These are the only particles substantially lighter than 1 TeV, besides the Bino-like $\tilde{\chi}^0_{3}$ and the RH sleptons, in particular the RH stau.
The squarks, gluino, Wino and LH sleptons are all in the multi-TeV range. 
Model B illustrates a solution with a light Bino and thus a Bino-like neutralino NLSP, as well as the
light RH stau and sleptons.
Model C is an example of a $\tilde{\chi}^0_{1}$ NLSP that is a substantial mixture of Higgsino and Bino.
Model D illustrates the corner of the parameter space with a stau NLSP (and mostly-Higgsino NNLSP). 
Finally, Model E shows the example of a spectrum corresponding to a low mediation scale, hence with a fast decaying NLSP, unlike the three previous cases.

The three definitions of fine tuning are also shown. From this comparison, we conclude that $\Delta_{\rm BG} \approx \Delta_{\rm EW} \ll \Delta_{\rm HS}$, which justifies our choice of basing our considerations mainly on $\Delta_{\rm EW}$. 

\section{Collider searches}
\label{sec:collider}
As we have shown above, our models feature multi-TeV colored superpartners. 
As a consequence, gauge mediation with non-unified messenger sectors can naturally accommodate at a low fine tuning price the experimental information of the 8 TeV run of the LHC: a 125 GeV Higgs, and no sign of supersymmetric particles. 

In this section, we discuss possible collider tests of the low-tuned solutions we found. 
We start by outlining the general features of the spectra of our models.
\begin{itemize}
\item 
The LSP is the gravitino, with mass given in terms of the SUSY-breaking F-term and the reduced Planck mass 
($M_P = 2.4\times 10^{18}$~GeV), see e.g.~\cite{martin}
\begin{equation}
\label{eq:mgr}
m_{\tilde G}= \frac{F}{\sqrt{3} M_P} = \frac{\Lambda \times M}{\sqrt{3} M_P}.
\end{equation}
Hence, $m_{\tilde G}$ can be in the range $\mathcal{O}(100)$ eV -- $\mathcal{O}(10)$~GeV, but we find that 
the fine-tuning
favours larger mediation scales and thus heavier gravitinos, cf.~Fig.\ref{fig:spectrum3} (right).
\item The NLSP is typically a sub-TeV Higgsino-like 
$\tilde{\chi}_1^0$ ($300~{\rm GeV}\lesssim\mu\lesssim 500$~GeV for $\Delta_{\rm EW}<50$), thus accompanied
by Higgsino-like $\tilde{\chi}_2^0$ and $\tilde{\chi}_1^\pm$, which are only $\mathcal{O}(10)$~GeV 
heavier than the NLSP.
Rather light Bino and RH sleptons are also possible, hence a corner of the parameter space features
$\tilde{\tau}_1$ as the NLSP, cf.~Fig.\ref{fig:spectrum2} (left), or alternatively a mostly Bino NLSP or
a mixed Bino-Higgsino NLSP.\\
A Higgsino NLSP can decay into $Z$ or $h$ and gravitino, while the decay of $\tilde{\chi}_1^0$ 
to photon is only possible in presence of a sizeable gaugino component. 
The neutralino decay widhts read \cite{Ambrosanio:1996jn,Dimopoulos:1996yq}:
\begin{align}
\label{eq:gammachi}
\Gamma ( \tilde{\chi}^0_1 \to \tilde{G}\, Z)  & \simeq\frac{m_{\tilde{\chi}^0_1}^5}{48 \pi\, m_{\tilde G}^2 M_{P}^2} 
\left[\left|N_{12} \cos\theta_W - N_{11}\sin\theta_W \right|^2 + \frac{1}{2}\left|N_{13} \cos\beta - N_{14}\sin\beta \right|^2\right]
\nonumber \\
& \times \left(1-\frac{m_{Z}^2}{m_{\tilde{\chi}^0_1}^2}  \right)^4, \\
\Gamma ( \tilde{\chi}^0_1 \to \tilde{G}\, h )  & \simeq\frac{m_{\tilde{\chi}^0_1}^5}{96 \pi\, m_{\tilde G}^2 M_{P}^2} 
\left|N_{13} \cos\beta + N_{14}\sin\beta \right|^2
\left(1-\frac{m_{h}^2}{m_{\tilde{\chi}^0_1}^2}  \right)^4, \\
\Gamma ( \tilde{\chi}^0_1 \to \tilde{G}\, \gamma )  & \simeq\frac{m_{\tilde{\chi}^0_1}^5}{48 \pi\, m_{\tilde G}^2 M_{P}^2} 
\left|N_{11} \cos\theta_W + N_{12}\sin\theta_W \right|^2, 
\label{eq:gammachigamma}
\end{align}
where $ N_{1k}$ are the gaugino/Higgsino components of the lightest neutralino, according to the definition:
$\tilde{\chi}_1^0 = N_{11} \tilde{B} + N_{12} \tilde{W}^0+ N_{13} \tilde{H}_d^0+ N_{14} \tilde{H}_u^0$.
In particular, in the pure Higgsino limit 
$|N_{11}|\simeq|N_{12}|\simeq0$ and $|N_{13}|\simeq|N_{14}|\simeq \sqrt{2}/2$.\\
A stau NLSP decays into $\tau$ and gravitino, see e.g.~\cite{martin}:
\begin{align}
\Gamma ( \tilde{\tau}_1 \to \tilde{G}\, \tau )& \simeq\frac{m_{\tilde{\tau}_1}^5}{48 \pi\, m_{\tilde G}^2 M_{P}^2},
\label{eq:gammatau}
\end{align}
As we can see, the NLSP coupling to the gravitino (and thus its decay rate) is inversely proportional to the gravitino mass itself and thus to the mediation scale $M$, cf.~Eq.~(\ref{eq:mgr}). 
Hence fine tuning prefers a long-lived NLSP. However, we find some solutions with a fast-decaying Higgsino NLSP, as discussed in the previous section. 
\item
The rest of the spectrum is in the multi-TeV mass range, in particular the colored superpartners.
As shown in Fig.\ref{fig:spectrum1} (left), the lightest stop mass is 
\begin{equation}
2~{\rm TeV}\lesssim m_{\tilde{t}_1} \lesssim 2.5~(5)~{\rm TeV}
\quad{\rm for}~\Delta_{\rm EW}\lesssim 50\,(100), 
\label{eq:stoprange}
\end{equation}
while the gluino mass is
\begin{equation}
3~{\rm TeV}\lesssim m_{\tilde{g}} \lesssim 5~(7)~{\rm TeV}
\quad{\rm for}~\Delta_{\rm EW}\lesssim 50\,(100).
\label{eq:gluinorange}
\end{equation}
The first and second generation squark masses are in a similar range as the gluino.
The LH ones are about 30\% heavier than the RH ones for $\Delta_{\rm EW}<50$, see 
Fig.\ref{fig:spectrum3} (left).
\end{itemize}
The above described features make the non-unified messenger models very difficult to test at the LHC, besides 
in some corners of the parameter space. In the following, we review the present status of the LHC searches and
the prospects of the LHC and future colliders for different relevant NLSP kinds and production modes. 
\vspace{-0.3cm}
\paragraph{The Long-lived Higgsino-like $\tilde{\chi}_1^0$ NLSP.} 
As we discussed at length, besides the gravitino LSP, the only particles required to lie well below the TeV are Higgsinos: $\tilde{\chi}_1^0$  (typically the NLSP), $\tilde{\chi}_2^0$ and $\tilde{\chi}^\pm_1$, both quasi-degenerate with $\tilde{\chi}_1^0$  (i.e.~with mass splittings at the percent level). Moreover, requiring low fine-tuning preferably selects
$M\gtrsim 10^{13}$~GeV, hence $\tilde{\chi}_1^0$ is typically long lived enough to escape the detector unseen.
This configuration resembles gravity mediation scenarios with a Higgsino NLSP and is among the most challenging to test at the LHC. 
In fact, the major production modes, i.e.~$\tilde{\chi}_1^0 \tilde{\chi}_2^0$, $\tilde{\chi}_{1,2}^0 \tilde{\chi}_1^\pm$, and $\tilde{\chi}_1^+ \tilde{\chi}_1^-$, lead to missing energy and very soft jets and/or leptons from the decays of $\tilde{\chi}_2^0$ and $\tilde{\chi}^\pm_1$ to the NLSP and off-shell $Z$ and $W$ bosons. 
Models A and C in Table~\ref{tab:benchmarks} provide examples of models of the above described kind.\\
Several studies in the literature have shown the limited LHC potential of testing models of this kind 
through Higgsino production, cf.~\cite{Han:2013usa,Gori:2013ala,Schwaller:2013baa,Baer:2014cua,Han:2014kaa}. 
Searches for missing transverse momentum ($\slashed{E}_T$) recoiling a single energetic jet (mono-$j$) 
or a single photon from initial state radiation will have no sensitivity to Higgsinos even at the future high-luminosity run of the LHC,
because the signal-to-background ratios are typically at the 1\% level~\cite{Han:2013usa,Baer:2014cua}.
An increased sensitivity is expected by selecting events with mono-$j$ and soft leptons. 
However, the estimated reach does not exceed $\approx 200$~GeV for the Higgsino mass 
after 300/fb of collected data at $\sqrt{s}=14$~TeV \cite{Han:2014kaa,Baer:2014kya}. 
This prospected sensitivity is below the Higgsino mass range of our models: 
$300~{\rm GeV}\lesssim \mu \lesssim500~{\rm GeV}$ for $\Delta_{\rm EW}<50$. 
Therefore, we can conclude that LHC is not able to test our low-tuned models (e.g.~Model A) even
in the long run.\\
For what concerns the proposed future colliders, the most promising possibility seems to be a leptonic ($e^+e^-$) 
machine with
high centre-of-mass energy. For instance, the International Linear Collider (ILC) operating at $\sqrt{s}=1$~TeV could
probe our models with $\Delta_{\rm EW}<50$, i.e., $\mu\lesssim 500$~GeV. 
Moreover, the ILC will be able to perform high precision measurements of the higgsino mass scale 
and associated mass gaps~\cite{Baer:2014yta}.
\vspace{-0.3cm}
\paragraph{The promptly decaying  Higgsino-like $\tilde{\chi}_1^0$ NLSP.}
A number of solutions feature a relatively low tuning, $\Delta_{\rm EW}\lesssim 100$, and a low mediation scale,
see Fig.~\ref{fig:spectrum3} (right). This case is characterized by a short-lived (neutralino) NLSP. An instance of 
such a setup is model E of Table~\ref{tab:benchmarks}: the resulting $\tilde{\chi}_1^0$ decay length is $c\tau_{\tilde \chi} = 0.1$ mm. Searches for promptly
decaying Higgsinos are therefore sensitive to this corner of the parameter space. 
As we have seen above, a Higgsino-like $\tilde{\chi}_1^0$ can decay to $h$ or $Z$ and $\slashed{E}_T$. 
These decay modes have been widely discussed in the literature as a signal of gauge mediation with a light gravitino \cite{Ambrosanio:1996jn,Dimopoulos:1996yq,Meade:2009qv,Ruderman:2011vv,Kats:2011qh}.
The CMS has published a search for neutralino and chargino production based on the data set of the $\sqrt{s}=8$~TeV
run \cite{Khachatryan:2014mma}. They interpret their results precisely in terms of production of promptly decaying Higgsinos in Gauge Mediation, setting a bound on the Higgsino mass as a function of 
${\rm BR} ( \tilde{\chi}^0_1 \to \tilde{G}\, h )$.  
For an almost pure Higgsino and moderate to large values of $\tan\beta$ (as it is the case of model E), the two decay modes slightly differ only by the phase space, typically giving ${\rm BR} ( \tilde{\chi}^0_1 \to \tilde{G}\, h ) = 40\div50 \%$ for the NSLP mass range of our models. The CMS limit then reads $m_{\tilde{\chi}_1^0} \gtrsim 300\div350$~GeV \cite{Khachatryan:2014mma}. \\
The total Higgsino production cross section for $pp$ collisions at 13 TeV is rather high: $\approx 34\,(15)$~fb for
$\mu = 500\,(600)$~GeV \cite{LHCxsec}. Hence, there are good prospects to test our low-tuned models 
with promptly decaying NLSP at the LHC.
\vspace{-0.3cm}
\paragraph{The long-lived Bino-like $\tilde{\chi}_1^0$ NLSP.}
Models of this kind (an example is model B in Table~\ref{tab:benchmarks}) can be tested through searches for electroweak production of Higgsinos, i.e., heavier neutralinos and chargino, and RH sleptons. In the former case, the most sensitive mode
is $pp\to \tilde{\chi}_{2,3}^0 \tilde{\chi}_1^\pm$ followed by $\tilde{\chi}_{2,3}^0 \to Z \tilde{\chi}_{1}^0$ and 
$\tilde{\chi}_{1}^\pm \to W^\pm \tilde{\chi}_{1}^0$ with the gauge bosons decaying leptonically. The direct slepton production leads
 to $\tilde{\ell}_R^+ \tilde{\ell}_R^- \to \ell^+ \ell^- \tilde{\chi}_{1}^0\tilde{\chi}_{1}^0$. Thus, searches for leptons and missing energy
 \cite{Aad:2014nua,Aad:2014vma,Khachatryan:2014qwa} are sensitive to this scenario. 
The present limits on RH sleptons are up to 250~GeV, only for a light Bino, $m_{\tilde{\chi}_1^0} \lesssim 100$~GeV.  
A similar bound on the Higgsino mass was obtained reinterpreting the searches for neutralinos and charginos decaying into $WZ$
in terms of Higgsino production \cite{Calibbi:2014lga}. Again, this bound can be evaded for neutralinos heavier than about 90 GeV.
However, the CMS prospects for the 14~TeV run with 300/fb \cite{CMSprosp} show that the $WZ$ channel can lead to a discovery
of (Wino-like) charginos-neutralinos up to 600~GeV, even for a Bino as heavy as 300~GeV. We can therefore conclude that the corner of our parameter space with a Bino-like NLSP have chances to be eventually tested at the LHC.
\vspace{-0.3cm}
\paragraph{The long-lived stau NLSP.}
Another class of relatively low-tuned solutions is characterized  by a RH stau as the NLSP. An example is 
 model D of Table~\ref{tab:benchmarks}, with $m_{\tilde{\tau}_1} = 533$~GeV and $c\tau_{\tilde \tau} = 2.8\times 10^6$ m.
 Such a long-lived charged NLSP releases energy throughout all layers of the detector due to its electromagnetic interactions with the material, hence it is reconstructed as a charged track, just like a muon. 
Searches for long-lived charged particles have been performed by both LHC experiments 
employing the 8~TeV run data \cite{Chatrchyan:2013oca,TheATLAScollaboration:2013qha}
and analysis based on the first 13 TeV collisions has been recently published by CMS \cite{CMS:2015kdx}.
The most stringent limit to date can be extracted by the 8~TeV CMS search \cite{Chatrchyan:2013oca}: interpreted
in terms of direct stau production, it reads $m_{\tilde{\tau}_1} \gtrsim 339$~GeV. Furthermore, this bound can
greatly increase if the NLSP is indirectly produced from cascade decays of heavier particles, such as Higgsinos in our case. For a recent discussion and a combination with other possibly relevant searches (e.g.~for disappearing tracks), we
 refer to \cite{Evans:2016zau}.
Since naturalness considerations (i.e.~$\Delta_{\rm EW}\lesssim 100$) imply for the stau NLSP scenario $m_{\tilde{\tau}_1} <  \mu \lesssim 650$~GeV, we expect that this region of our parameter space can be fully covered by 
the 13/14~TeV run of the LHC.
\vspace{-0.3cm}
\paragraph{Production of colored superpartners.}
As we have seen, even the models with lowest tuning are charachterized by super-heavy strongly-interacting particles,
cf.~Eqs~(\ref{eq:stoprange},\ref{eq:gluinorange}). Thus the colored spectrum is beyond the reach of the LHC, given the future limit e.g.~on the stop mass, as estimated by the collaborations themselves: 
 $m_{\tilde{t}_1} \gtrsim 950$~GeV for $\sqrt{s}=14$~TeV and 300/fb \cite{CMSprosp}, 
 $m_{\tilde{t}_1} \gtrsim 1450$~GeV for $\sqrt{s}=14$~TeV and 3000/fb \cite{ATLASprosp}.
In general, all studies limit the reach in the gluino and first generation squark masses to $2\div 3$~TeV, 
see e.g.~\cite{Cohen:2013xda}.
It is then clear that only a future high-energy hadronic collider can fully test 
the natural configurations of non-unified gauge mediation that we found (cf.~Fig.~\ref{fig:spectrum1}, left), 
and the characteristic features of the spectrum, in particular the predicted hierarchy of LH and RH squarks (see Fig.~\ref{fig:spectrum3}, right).  
The reach of a $\sqrt{s}=100$~TeV $pp$ collider is indeed assessed to be at the order of 10$\div$15~TeV for the gluino/squark masses \cite{Cohen:2013xda,Ellis:2015xba}.

\section{Gravitino Dark Matter}
\label{sec:DM}
\begin{table}[t]
{\small
\begin{center}
\begin{tabular}{@{} |c||c|c|c|c|c| @{}}
\hline    \hline    
     & A & B & C & D & E\\ 
\hline    
 $m_{\tilde G}$  & 640 MeV & 25 MeV &  26 GeV & 19 MeV  & 175 eV\\ 
 $T_{RH}^*$   [GeV] & $4.6\times 10^5$ & $2.2\times 10^4$ & $2.4\times 10^7$ & $9\times 10^3$  & - \\ 
 \hline
 NLSP    &  $\tilde{\chi}^0_1$ & $\tilde{\chi}^0_1$  & $\tilde{\chi}^0_1$  & $\tilde{\tau}_1$ & $\tilde{\chi}^0_1$ \\ 
 \hline
 $m_{\rm NLSP}$  [GeV]  & 320 &  455 &  571 & 533  & 621 \\ 
 $\tau_{\rm NLSP}$ [s]  & 144  &  0.039 & $1.3\times 10^4$ & $9.4\times 10^{-3}$ &  $3.7\times 10^{-13}$  \\ 
 $\Omega_{\rm NLSP} h^2$ & $2.1\times10^{-2}$ &  1.9 & $0.12$  & $8.5\times10^{-2}$ & $7.5\times10^{-2}$ \\ 
 \hline  \hline    
  \end{tabular}
 \end{center}
 }
\caption{Quantities relevant to gravitino cosmology for the models of Table \ref{tab:benchmarks}. \label{tab:DM}}
\end{table}
In this section, we comment about the viability of the gravitino LSP as a Dark Matter (DM) candidate in our models. 
A systematic discussion is beyond the scope of the present study. Here, we just focus 
on the five examples displayed in Table~\ref{tab:benchmarks}, in order to illustrate 
possible issues and further constraints arising from the requirement of a consistent gravitino cosmology, e.g.~from Big Bang Nucleosynthesis (BBN). In fact, as we have seen in the previous section, 
the NLSP is typically long-lived in our framework, so that it undergoes late decays that can spoil the successful predictions of BBN.
This leads to bounds that apply for $\tau_{\rm NLSP} > 10^{-2}$ s and we will read from \cite{Jedamzik:2006xz,CahillRowley:2012cb}. 
As we discussed, typical non-unified GM models with low fine tuning feature a neutralino NLSP, but a stau NLSP is also possible in our framework.
For an early review on gravitino cosmology we refer to \cite{Giudice:1998bp}.
Gravitino DM with a generic neutralino NLSP has been discussed in \cite{Covi:2009bk}.  
A review of gravitino cosmology with stau NLSP can be found in \cite{Steffen:2006hw}.
For a recent studies within the phenomenological MSSM, see also \cite{CahillRowley:2012cb,Arvey:2015nra}. 

Gravitinos can be produced in the early Universe both thermally, from scattering of SUSY particles \cite{Bolz:2000fu,Pradler:2006qh}, and non-thermally, from decays of the NLSP after freeze-out \cite{Covi:1999ty,Feng:2003uy}. 
Thermal production always dominate in our scenarios with gravitino cold DM. 
This mode depends on the reheating temperature $T_{RH}$ such that an upper bound
on $T_{RH}$ is obtained by requiring that the relic density of the gravitino $\Omega_{\tilde G}$ does not exceed the observed DM abundance $\Omega_{\rm DM}$. We compute $\Omega_{\tilde G}$ for our benchmark models by means of the analytical
expressions in \cite{Pradler:2006qh,Steffen:2006hw}.
In Table~\ref{tab:DM}, we display all information we need: 
the gravitino mass, the value $T_{RH}^*$ for the reheating temperature that saturates the above mentioned bound (i.e.~that corresponds to $\Omega_{\tilde G} h^2 = \Omega_{\rm DM} h^2 \simeq 0.12$ \cite{Ade:2013zuv}), as well as the NLSP mass, life-time, and freeze-out energy density $\Omega_{\rm NSLP} h^2$. 
The last quantities are computed by means of {\tt ISAJET} and {\tt micrOMEGAs}, while the NLSP life-time employing Eqs.~(\ref{eq:gammachi}\,-\ref{eq:gammatau}).
\vspace{-0.3cm}
\paragraph{Model A.}
As we have already discussed, this model is representative of the solutions with lowest $\Delta_{\rm EW}$ that we found in our numerical analysis. Given that high mediation scales are preferred by fine-tuning, these solutions typically have $m_{\tilde G}\gtrsim \mathcal{O}(0.1\div 1)$~GeV (in this case 640 MeV). 
Also, a Higgsino-like $\tilde{\chi}^0_1$ NLSP is selected. 
From the properties of the gravitino and the NLSP displayed in Table~\ref{tab:DM}, we also learn the following. 
The NLSP decays during the BBN epoch ($\tau_{\rm NLSP}=144$ s), mostly to baryons 
(following $\tilde{\chi}^0_1\to \tilde{G}\,Z/h$). 
However, given the large annihilation cross section of Higgsinos, the NLSP is diluted enough 
($\Omega_{\rm NLSP} h^2=2.1\times10^{-2}$) that this model marginally evades
the bounds displayed in  \cite{Jedamzik:2006xz,CahillRowley:2012cb}. 
This is not the case of models with heavier gravitinos, i.e.~larger 
$\tau_{\rm NLSP}$. Hence, BBN constraints can exclude part of our solutions with low tuning, as far as a standard
thermal history of the universe is assumed, no extra dilution mechanism is at work, etc. 
The gravitino of model A is a viable cold DM candidate and accounts for the observed relic density if 
$T_{RH} = \mathcal{O}(10^5)$~GeV. This value is about three orders of magnitude below the lower bound to 
$T_{RH}$ posed by leptogenesis (for a recent review see \cite{Fong:2013wr}). Hence model A would require a low-temperature
baryogenesis mechanism or, again, extra entropy production like e.g.~in \cite{Hasenkamp:2010if}.
\vspace{-0.3cm}
\paragraph{Model B.}
The peculiar feature of model B is that the NLSP is mostly Bino. This translates to a larger energy density of the NLSP at freeze-out, 
 $\Omega_{\rm NLSP} h^2 = 1.9$. Still, this gives a negligible non-thermal contribution to the gravitino relic density, given
that the energy density is suppressed by the small gravitino mass:  
$\Omega^{\rm NT}_{\tilde G} =  \Omega_{\rm NLSP} \times (m_{\tilde G}/ m_{\rm NLSP})$ \cite{Covi:1999ty,Feng:2003uy}.
Furthermore, again in spite of the large $\Omega_{\rm NLSP}$, 
model B is not excluded by the BBN constraints of \cite{Jedamzik:2006xz,CahillRowley:2012cb}, due to the 
relatively fast neutralino decay, as well as the reduced baryonic decay width. In fact BR$(\tilde{\chi}^0_1 \to {\rm baryons})\simeq$
13\%, since $\tilde{\chi}^0_1\to \tilde{G}\,\gamma$ is the dominant decay mode of a Bino NLSP, cf.~Eq.~(\ref{eq:gammachi},\,\ref{eq:gammachigamma}).
Finally, we notice that also in this case a relatively low reheating temperature, $\lesssim\mathcal{O}(10)$~TeV, is needed to avoid DM overproduction.
\vspace{-0.3cm}
\paragraph{Model C.}
This model features a mixed Higgsino-Bino NLSP. As we can see from Table~\ref{tab:DM}, the gravitino is rather heavy (26 GeV) and thus the NLSP life time is long, $\mathcal{O}(10^4)$\,s, compared to the BBN time scale. As a consequence, this is an example of model excluded by the BBN constraints of \cite{Jedamzik:2006xz,CahillRowley:2012cb}, unless some non-standard mechanism intervenes to reduce the NLSP energy density by at least three orders of magnitude.
\vspace{-0.3cm}
\paragraph{Model D.}
This model is our example of a stau NLSP.  The BBN constraints for a charged NLSP can be read for instance in \cite{CahillRowley:2012cb}. They are fulfilled in the case of model D, due to a fast stau decay, $\mathcal{O}(10^{-2})$\,s. 
Therefore, this provides another example of a viable model of gravitino DM, as far as $T_{\rm RH}$ is about 10~TeV, similarly to model B. 
\vspace{-0.3cm}
\paragraph{Model E.}
This model illustrates the case a neutralino NLSP promptly decaying at colliders, which requires a very low gravitino mass. 
As a consequence, the gravitino of model E is not a cold DM candidate and its relic density must be strongly suppressed by some non-standard mechanism, in order to evade the stringent cosmological constraints on hot and warm DM. An example of such mechanism can be found in \cite{Baltz:2001rq}. Although this model does not provide a DM candidate, the BBN constraints 
are trivially satisfied due to the tiny life time of the NLSP.

\section{Models of messengers}
\label{sec:models}
To cancel the gauge anomalies, we consider the messengers as chiral superfields in vector-like representations of the 
SM gauge group.
Furthermore, in order to preserve the gauge coupling unification, we introduce additional vector-like particles
at the messenger scale, which however do not couple to the SUSY breaking sector and only act as ``spectators''.
The SM and supersymmetric SMs with vector-like particles have been studied extensively previously,
for instance, see Refs.~\cite{Li:2010hi, LMN, Jiang:2006hf, Barger:2006fm, Barger:2007qb, Jiang:2008yf,
Jiang:2009za, Li:2009cy}. 
In particular, the one-loop beta function equivalent relations among the different particle sets have been
studied as well~\cite{LMN, Barger:2007qb}.
Thus, we will present one set of the messenger fields and vector-like particles for each benchmark point,
and all the other sets can be obtained via the one-loop beta function equivalent relations~\cite{LMN, Barger:2007qb}.

To be concrete, at the messenger scale, we introduce the messenger fields and vector-like particles,
whose contributions to the one-loop beta function coefficients 
are denoted as $(b_1^M,~b_2^M,~b_3^M)$ and $(b_1^V,~b_2^V,~b_3^V)$,
respectively. To preserve the gauge coupling unification, 
we require
\begin{align}
b_1^{M+V} ~=~b_2^{M+V} ~=~b_3^{M+ V} \equiv b^{M+ V}~.~
\end{align}
where $b_i^{M+V}\equiv b_i^M+b_i^V$. Other relations among the $b_i^{M+V}$ could also preserve unification, 
see \cite{Calibbi:2009cp}. We neglect these possibilities for simplicity.   
As in the previous studies, we only introduce
the messenger and vector-like particles whose quantum numbers are the same as those
of the SM fermions and their Hermitian conjugates, and
the $SU(5)$ adjoint particles. Their quantum numbers under 
$SU(3) \times SU(2) \times U(1)$ and their
contributions to one-loop beta functions, $\Delta b \equiv (\Delta
b_1, \Delta b_2, \Delta b_3)$ as complete supermultiplets are
\begin{eqnarray}
& & XQ + {\overline{XQ}} = {\mathbf{(3, 2, {1\over 6}) + ({\bar 3}, 2,
-{1\over 6})}}\,, \quad \Delta b =({1\over 5}, 3, 2)\,;\\ 
& & XU + {\overline{XU}} = {\mathbf{ ({3},
1, {2\over 3}) + ({\bar 3},  1, -{2\over 3})}}\,, \quad \Delta b =
({8\over 5}, 0, 1)\,;\\ 
& & XD + {\overline{XD}} = {\mathbf{ ({3},
1, -{1\over 3}) + ({\bar 3},  1, {1\over 3})}}\,, \quad \Delta b =
({2\over 5}, 0, 1)\,;\\  
& & XL + {\overline{XL}} = {\mathbf{(1,  2, {1\over 2}) + ({1},  2,
-{1\over 2})}}\,, \quad \Delta b = ({3\over 5}, 1, 0)\,;\\ 
& & XE + {\overline{XE}} = {\mathbf{({1},  1, {1}) + ({1},  1,
-{1})}}\,, \quad \Delta b = ({6\over 5}, 0, 0)\,;\\ 
& & XG = {\mathbf{({8}, 1, 0)}}\,, \quad \Delta b = (0, 0, 3)\,;\\ 
& & XW = {\mathbf{({1}, 3, 0)}}\,, \quad \Delta b = (0, 2, 0)\,;\\
& & XY + {\overline{XY}} = {\mathbf{(3, 2, -{5\over 6}) + ({\bar 3}, 2,
{5\over 6})}}\,, \quad \Delta b =(5, 3, 2)\,.\,
\end{eqnarray}

\begin{table}[t]
{\small
\begin{center}
\begin{tabular}{@{} |c||c|c|c| @{}}
\hline    \hline    
     & A/C & B/D & E\\ 
\hline    
Messengers & $(XQ,~{\overline{XQ}})$, $3(XD,~{\overline{XD}})$,  &  $2(XQ,~{\overline{XQ}})$, $2 XW$, 
&  $2(XD,~{\overline{XD}})$,   \\
& $4(XL,~{\overline{XL}})$, $3 XW$  & $n(XE,~{\overline{XE}})$ & $2(XL,~{\overline{XL}})$, $XW$\\

 \hline
 $(b_1^M,~b_2^M,~b_3^M)$  & $(\frac{19}{5}, ~13, ~5)$ & $(\frac{2+6n}{5}, ~10, ~4)$  & $(2, ~4, ~2)$ \\ 
 \hline
 VLPs & $3(XU,~{\overline{XU}})$, $5(XD,~{\overline{XD}})$, &  $(XU,~{\overline{XU}})$, $2(XD,~{\overline{XD}})$,
 &  $(XD,~{\overline{XD}})$, \\
 & $2(XE,~{\overline{XE}})$ & $(6-n)(XE,~{\overline{XE}})$, $XG$ & $(XU,~{\overline{XU}})$\\
\hline
$(b_1^V,~b_2^V,~b_3^V)$  & $(\frac{46}{5}, ~0, ~8)$  &  $(\frac{48-6n}{5}, ~0, ~6)$  &
$(2, ~0, ~2)$
\\ 
 \hline
$b^{M+V}$  & $13$  & $10$ &  $4$ \\
 \hline  \hline    
  \end{tabular}
 \end{center}
 }
\caption{ The messenger fields, Vector-Like Particles (VLPs) and their one-loop beta functions
in our Models. Here, $n=0, ~1, ..., ~6$.\label{tab:MPB}}
\end{table}
We present the messenger fields, vector-like particles and their one-loop beta functions
for our benchmark models in Table \ref{tab:MPB}.

\section{Conclusions}
\label{sec:conclusions}

We have studied the fine tuning of models of gauge-mediated SUSY breaking, assuming general sets of messenger fields. We focused in particular on messengers not belonging to complete representations of grand-unified gauge groups, thus departing from the framework of minimal GM. 
The outcome of this study can be summarized as follows:
\begin{itemize}
\item Models with non-unified messenger sectors can improve the fine tuning up to one order of magnitude with respect of minimal GM. In particular, we found solutions featuring an electroweak fine tuning $\Delta_{\rm EW}$ as low as 40$\div$50,
as shown in Figs.~\ref{fig:mh-FT} and \ref{fig:comp}.
\item This can occur if the messenger sector is such that the Wino mass is substantially larger than the gluino mass at the mediation scale, due to a compensating effect between gauge and Yukawa contributions in the ${\tilde m}^2_{H_u}$ RGE, cf.~Eq.~(\ref{eq:rge}).
\item Although the fine tuning is low, the spectrum of our models lies in the multi-TeV range, due to the requirement of heavy stops to raise the Higgs mass to the observed value, as well as heavy Wino for the compensating effect in the running ${\tilde m}^2_{H_u}$.
As a consequence, the only possibly light states are the Higgsinos and the singlets under both 
$SU(3)$ and $SU(2)$: Bino and RH sleptons. 
The absence of signs of SUSY at the LHC is therefore a natural consequence of our framework.
\item The LSP is a light gravitino, while the NLSP is typically an Higgsino-like neutralino, although corners of the parameter space can feature either a Bino or a RH stau NLSP. Fine tuning favours solutions with a large mediation scale, thus with a NLSP that appears as a long-lived particle at colliders. These features challenge LHC searches for SUSY particles at the LHC, as discussed in detail in section \ref{sec:collider}. Although some corners of the parameter space will be tested at the 14 TeV run (those with long-lived stau NLSP, Bino-like NLSP, or a promptly decaying neutralino NLSP), the typical low-tuned models we found shall await a future leptonic or hadronic machine to be probed. 
\item As shown in section \ref{sec:DM}, the gravitino LSP is produced mainly by thermal scattering in the early universe and can be a consistent candidate for cold dark matter, provided that its relic density is diluted by means of a rather low reheating temperature,
or an alternative non-standard mechanism of e.g.~enthropy production. 
The stringent BBN bounds on NLSP decays can be evaded unless the mediation scale is very high.
\item As shown in section \ref{sec:models}, sets of messengers in incomplete multiplets giving low fine tuning can be easily built, while gauge coupling unification can be preserved by introducing additional `spectator' vector-like fields.
\end{itemize}

\section*{Acknowledgments}

This work is supported in part by the
Natural Science Foundation of China grant numbers 11135003, 11275246, and 11475238 (TL).


\begin{thebibliography}{999}


\bibitem{Aad:2012tfa}
  G.~Aad {\it et al.} [ATLAS Collaboration],
  Phys.\ Lett.\ B {\bf 716} (2012) 1
  [arXiv:1207.7214 [hep-ex]].

\bibitem{Chatrchyan:2012xdj}
  S.~Chatrchyan {\it et al.} [CMS Collaboration],
  Phys.\ Lett.\ B {\bf 716} (2012) 30
  [arXiv:1207.7235 [hep-ex]].


\bibitem{Craig:2013cxa} 
  N.~Craig,
  arXiv:1309.0528 [hep-ph].
  
\bibitem{Arvanitaki:2013yja}
  A.~Arvanitaki, M.~Baryakhtar, X.~Huang, K.~van Tilburg and G.~Villadoro,
  JHEP {\bf 1403} (2014) 022
  [arXiv:1309.3568 [hep-ph]].
 
\bibitem{Giudice:1998bp}
  G.~F.~Giudice and R.~Rattazzi,
  Phys.\ Rept.\  {\bf 322} (1999) 419
  [hep-ph/9801271].

\bibitem{Gogoladze:2009bd} 
  I.~Gogoladze, M.~U.~Rehman and Q.~Shafi,
  Phys.\ Rev.\ D {\bf 80}, 105002 (2009)
  [arXiv:0907.0728 [hep-ph]].
  
\bibitem{Horton:2009ed}
  D.~Horton and G.~G.~Ross,
  Nucl.\ Phys.\ B {\bf 830} (2010) 221
  [arXiv:0908.0857 [hep-ph]].

\bibitem{Antusch:2012gv}
  S.~Antusch, L.~Calibbi, V.~Maurer, M.~Monaco and M.~Spinrath,
  JHEP {\bf 1301} (2013) 187
  [arXiv:1207.7236].

\bibitem{Gogoladze:2012yf} 
  I.~Gogoladze, F.~Nasir and Q.~Shafi,
  Int.\ J.\ Mod.\ Phys.\ A {\bf 28}, 1350046 (2013)
  [arXiv:1212.2593 [hep-ph]].

\bibitem{Gogoladze:2013wva} 
  I.~Gogoladze, F.~Nasir and Q.~Shafi,
  JHEP {\bf 1311}, 173 (2013)
  [arXiv:1306.5699 [hep-ph]].

\bibitem{Kaminska:2013mya} 
  A.~Kaminska, G.~G.~Ross and K.~Schmidt-Hoberg,
  JHEP {\bf 1311}, 209 (2013)
  [arXiv:1308.4168 [hep-ph]].


\bibitem{Martin:2013aha}
  S.~P.~Martin,
  Phys.\ Rev.\ D {\bf 89} (2014) 3,  035011
  [arXiv:1312.0582 [hep-ph]].
  
 \bibitem{Kowalska:2014hza}
  K.~Kowalska, L.~Roszkowski, E.~M.~Sessolo and S.~Trojanowski,
  JHEP {\bf 1404} (2014) 166
  [arXiv:1402.1328 [hep-ph]].

\bibitem{Li:2010hi}
  T.~Li and D.~V.~Nanopoulos,
  JHEP {\bf 1110} (2011) 090
  [arXiv:1005.3798 [hep-ph]].


\bibitem{Brummer:2012zc}
  F.~Brummer and W.~Buchmuller,
  JHEP {\bf 1205} (2012) 006
  [arXiv:1201.4338 [hep-ph]].



\bibitem{Brummer:2013yya}
  F.~Brümmer, M.~Ibe and T.~T.~Yanagida,
  Phys.\ Lett.\ B {\bf 726} (2013) 364
  [arXiv:1303.1622 [hep-ph]].



\bibitem{Bhattacharyya:2015vha}
  G.~Bhattacharyya, T.~T.~Yanagida and N.~Yokozaki,
  Phys.\ Lett.\ B {\bf 749} (2015) 82
  [arXiv:1506.05962 [hep-ph]].



\bibitem{Fukuda:2015pra}
  H.~Fukuda, H.~Murayama, T.~T.~Yanagida and N.~Yokozaki,
  Phys.\ Rev.\ D {\bf 92} (2015) no.5,  055032
  [arXiv:1508.00445 [hep-ph]].


\bibitem{Li:2014xqa} 
  T.~Li, D.~V.~Nanopoulos, S.~Raza and X.~C.~Wang,
  JHEP {\bf 1408}, 128 (2014)
  [arXiv:1406.5574 [hep-ph]].

\bibitem{Casas:2014eca}
  J.~A.~Casas, J.~M.~Moreno, S.~Robles, K.~Rolbiecki and B.~Zaldívar,
  JHEP {\bf 1506} (2015) 070
  [arXiv:1407.6966 [hep-ph]].


\bibitem{Li:2014dna} 
  T.~Li and S.~Raza,
  Phys.\ Rev.\ D {\bf 91}, no. 5, 055016 (2015)
  [arXiv:1409.3930 [hep-ph]].

\bibitem{Du:2015una}
  G.~Du, T.~Li, D.~V.~Nanopoulos and S.~Raza,
  Phys.\ Rev.\ D {\bf 92} (2015) no.2,  025038
  [arXiv:1502.06893 [hep-ph]].


\bibitem{Ding:2015vla}
  R.~Ding, T.~Li, L.~Wang and B.~Zhu,
  JHEP {\bf 10} (2015) 154
  [arXiv:1506.00359 [hep-ph]].

\bibitem{Ding:2015epa}
  R.~Ding, T.~Li, F.~Staub and B.~Zhu,
  arXiv:1510.01328 [hep-ph].

\bibitem{Li:2016ucz} 
  T.~Li, S.~Raza and K.~Wang,
  arXiv:1601.00178 [hep-ph].

\bibitem{Casas:2016xnl}
  J.~A.~Casas, J.~M.~Moreno, S.~Robles and K.~Rolbiecki,
  arXiv:1602.06892 [hep-ph].


\bibitem{Barbieri:1987fn}
  R.~Barbieri and G.~F.~Giudice,
  Nucl.\ Phys.\ B {\bf 306} (1988) 63.

\bibitem{Martin:1996zb}
  S.~P.~Martin,
  Phys.\ Rev.\ D {\bf 55} (1997) 3177
  [hep-ph/9608224].
  

\bibitem{Baer:2012up}
  H.~Baer, V.~Barger, P.~Huang, A.~Mustafayev and X.~Tata,
  Phys.\ Rev.\ Lett.\  {\bf 109} (2012) 161802
  [arXiv:1207.3343 [hep-ph]].
  
\bibitem{Baer:2012mv}
  H.~Baer, V.~Barger, P.~Huang, D.~Mickelson, A.~Mustafayev and X.~Tata,
  Phys.\ Rev.\ D {\bf 87} (2013) 3,  035017
  [arXiv:1210.3019 [hep-ph]].
  
\bibitem{Baer:2012cf} 
  H.~Baer, V.~Barger, P.~Huang, D.~Mickelson, A.~Mustafayev and X.~Tata,
  Phys.\ Rev.\ D {\bf 87}, no. 11, 115028 (2013)
  [arXiv:1212.2655 [hep-ph]].
  
\bibitem{Baer:2013gva}
  H.~Baer, V.~Barger and D.~Mickelson,
  Phys.\ Rev.\ D {\bf 88} (2013) 9,  095013
  [arXiv:1309.2984 [hep-ph]].

\bibitem{Mustafayev:2014lqa} 
  A.~Mustafayev and X.~Tata,
  Indian J.\ Phys.\  {\bf 88}, 991 (2014)
  [arXiv:1404.1386 [hep-ph]].


\bibitem{Baer:1999sp} 
  H.~Baer, F.~E.~Paige, S.~D.~Protopopescu and X.~Tata,
  hep-ph/0001086.

\bibitem{ATLAS:2014wva} 
  [ATLAS and CDF and CMS and D0 Collaborations],
  arXiv:1403.4427 [hep-ex].

\bibitem{Gogoladze:2015tfa} 
  I.~Gogoladze, A.~Mustafayev, Q.~Shafi and C.~S.~Un,
  Phys.\ Rev.\ D {\bf 91}, no. 9, 096005 (2015)
[arXiv:1501.07290 [hep-ph]].

 \bibitem{Ajaib:2012vc}                                                 
 M.~A.~Ajaib, I.~Gogoladze, F.~Nasir and Q.~Shafi,                                       
 Phys.\ Lett.\ B {\bf 713}, 462 (2012)            
 [arXiv:1204.2856 [hep-ph]].


\bibitem{Meade:2008wd}
  P.~Meade, N.~Seiberg and D.~Shih,
  Prog.\ Theor.\ Phys.\ Suppl.\  {\bf 177} (2009) 143
  [arXiv:0801.3278 [hep-ph]].

\bibitem{Knapen:2015qba}
  S.~Knapen, D.~Redigolo and D.~Shih,
  arXiv:1507.04364 [hep-ph].

\bibitem{Grajek:2013ola}
  P.~Grajek, A.~Mariotti and D.~Redigolo,
  JHEP {\bf 1307} (2013) 109
  [arXiv:1303.0870 [hep-ph]].


\bibitem{martin}
  S.~P.~Martin,
  Adv.\ Ser.\ Direct.\ High Energy Phys.\  {\bf 21} (2010) 1
   [Adv.\ Ser.\ Direct.\ High Energy Phys.\  {\bf 18} (1998) 1] 
  [hep-ph/9709356].


\bibitem{Han:2013usa}
  C.~Han, A.~Kobakhidze, N.~Liu, A.~Saavedra, L.~Wu and J.~M.~Yang,
  JHEP {\bf 1402} (2014) 049
  [arXiv:1310.4274 [hep-ph]].

\bibitem{Gori:2013ala}
  S.~Gori, S.~Jung and L.~T.~Wang,
  JHEP {\bf 1310} (2013) 191
  [arXiv:1307.5952 [hep-ph]].

\bibitem{Schwaller:2013baa}
  P.~Schwaller and J.~Zurita,
  JHEP {\bf 1403} (2014) 060
  [arXiv:1312.7350 [hep-ph]].

\bibitem{Baer:2014cua} 
  H.~Baer, A.~Mustafayev and X.~Tata,
  Phys.\ Rev.\ D {\bf 89}, no. 5, 055007 (2014)
  [arXiv:1401.1162 [hep-ph]].

\bibitem{Han:2014kaa}
  Z.~Han, G.~D.~Kribs, A.~Martin and A.~Menon,
  Phys.\ Rev.\ D {\bf 89} (2014) 7,  075007
  [arXiv:1401.1235 [hep-ph]].

\bibitem{Baer:2014kya} 
  H.~Baer, A.~Mustafayev and X.~Tata,
  Phys.\ Rev.\ D {\bf 90}, no. 11, 115007 (2014)
  [arXiv:1409.7058 [hep-ph]].

\bibitem{Baer:2014yta} 
  H.~Baer, V.~Barger, D.~Mickelson, A.~Mustafayev and X.~Tata,
  JHEP {\bf 1406}, 172 (2014)
  [arXiv:1404.7510 [hep-ph]].



\bibitem{Ambrosanio:1996jn}
  S.~Ambrosanio, G.~L.~Kane, G.~D.~Kribs, S.~P.~Martin and S.~Mrenna,
  Phys.\ Rev.\ D {\bf 54} (1996) 5395
  [hep-ph/9605398].

\bibitem{Dimopoulos:1996yq}
  S.~Dimopoulos, S.~D.~Thomas and J.~D.~Wells,
  Nucl.\ Phys.\ B {\bf 488} (1997) 39
  [hep-ph/9609434].


\bibitem{Meade:2009qv}
  P.~Meade, M.~Reece and D.~Shih,
  JHEP {\bf 1005} (2010) 105
  [arXiv:0911.4130 [hep-ph]].

\bibitem{Ruderman:2011vv}
  J.~T.~Ruderman and D.~Shih,
  JHEP {\bf 1208} (2012) 159
  [arXiv:1103.6083 [hep-ph]].

\bibitem{Kats:2011qh}
  Y.~Kats, P.~Meade, M.~Reece and D.~Shih,
  JHEP {\bf 1202} (2012) 115
  [arXiv:1110.6444 [hep-ph]].

\bibitem{Khachatryan:2014mma}
  V.~Khachatryan {\it et al.} [CMS Collaboration],
  Phys.\ Rev.\ D {\bf 90} (2014) 9,  092007
  [arXiv:1409.3168 [hep-ex]].

\bibitem{LHCxsec}
  M.~Kramer, A.~Kulesza, R.~van der Leeuw, M.~Mangano, S.~Padhi, T.~Plehn and X.~Portell,
  arXiv:1206.2892 [hep-ph];
  B.~Fuks, M.~Klasen, D.~R.~Lamprea and M.~Rothering,
  JHEP {\bf 1210} (2012) 081
  [arXiv:1207.2159 [hep-ph]];
  B.~Fuks, M.~Klasen, D.~R.~Lamprea and M.~Rothering,
  Eur.\ Phys.\ J.\ C {\bf 73} (2013) 2480
  [arXiv:1304.0790 [hep-ph]];
 and LHC SUSY Cross Section Working Group,
\url{https://twiki.cern.ch/twiki/bin/view/LHCPhysics/SUSYCrossSections}.



\bibitem{Aad:2014nua}
  G.~Aad {\it et al.} [ATLAS Collaboration],
  JHEP {\bf 1404} (2014) 169
  [arXiv:1402.7029 [hep-ex]].

\bibitem{Aad:2014vma}
  G.~Aad {\it et al.} [ATLAS Collaboration],
  JHEP {\bf 1405} (2014) 071
  [arXiv:1403.5294 [hep-ex]].

\bibitem{Khachatryan:2014qwa}
  V.~Khachatryan {\it et al.} [CMS Collaboration],
  Eur.\ Phys.\ J.\ C {\bf 74} (2014) 9,  3036
  [arXiv:1405.7570 [hep-ex]].

\bibitem{Calibbi:2014lga}
  L.~Calibbi, J.~M.~Lindert, T.~Ota and Y.~Takanishi,
  JHEP {\bf 1411} (2014) 106
  [arXiv:1410.5730 [hep-ph]].
  
\bibitem{CMSprosp}
  The CMS Collaboration,
  arXiv:1307.7135.


\bibitem{Chatrchyan:2013oca}
  S.~Chatrchyan {\it et al.} [CMS Collaboration],
  JHEP {\bf 1307} (2013) 122
  [arXiv:1305.0491 [hep-ex]].

\bibitem{TheATLAScollaboration:2013qha}
  The ATLAS Collaboration,
  ATLAS-CONF-2013-058.

\bibitem{CMS:2015kdx}
  The CMS Collaboration,
  CMS-PAS-EXO-15-010.
  
\bibitem{Evans:2016zau}
  J.~A.~Evans and J.~Shelton,
  arXiv:1601.01326 [hep-ph].



\bibitem{ATLASprosp}
  The ATLAS Collaboration,
   ATL-PHYS-PUB-2013-011.  

\bibitem{Cohen:2013xda}
  T.~Cohen, T.~Golling, M.~Hance, A.~Henrichs, K.~Howe, J.~Loyal, S.~Padhi and J.~G.~Wacker,
  JHEP {\bf 1404} (2014) 117
  [arXiv:1311.6480 [hep-ph]].
  
\bibitem{Ellis:2015xba}
  S.~A.~R.~Ellis and B.~Zheng,
  Phys.\ Rev.\ D {\bf 92} (2015) 7,  075034
  [arXiv:1506.02644 [hep-ph]].
  
  
   
\bibitem{Jedamzik:2006xz}
  K.~Jedamzik,
  Phys.\ Rev.\ D {\bf 74} (2006) 103509
  [hep-ph/0604251].
  
\bibitem{CahillRowley:2012cb}
  M.~W.~Cahill-Rowley, J.~L.~Hewett, S.~Hoeche, A.~Ismail and T.~G.~Rizzo,
  Eur.\ Phys.\ J.\ C {\bf 72} (2012) 2156
  [arXiv:1206.4321 [hep-ph]].
  
 \bibitem{Covi:2009bk}
  L.~Covi, J.~Hasenkamp, S.~Pokorski and J.~Roberts,
  JHEP {\bf 0911} (2009) 003
  [arXiv:0908.3399 [hep-ph]].
  
\bibitem{Steffen:2006hw}
  F.~D.~Steffen,
  JCAP {\bf 0609} (2006) 001
  [hep-ph/0605306].
  
  
\bibitem{Arvey:2015nra}
  A.~Arbey, M.~Battaglia, L.~Covi, J.~Hasenkamp and F.~Mahmoudi,
  Phys.\ Rev.\ D {\bf 92} (2015) 11,  115008
  [arXiv:1505.04595 [hep-ph]].
  
\bibitem{Bolz:2000fu}
  M.~Bolz, A.~Brandenburg and W.~Buchmuller,
  Nucl.\ Phys.\ B {\bf 606} (2001) 518
   [Nucl.\ Phys.\ B {\bf 790} (2008) 336]
  [hep-ph/0012052].
  

\bibitem{Pradler:2006qh}
  J.~Pradler and F.~D.~Steffen,
  Phys.\ Rev.\ D {\bf 75} (2007) 023509
  [hep-ph/0608344].
  
\bibitem{Covi:1999ty}
  L.~Covi, J.~E.~Kim and L.~Roszkowski,
  Phys.\ Rev.\ Lett.\  {\bf 82} (1999) 4180
  [hep-ph/9905212].
  
  
\bibitem{Feng:2003uy}
  J.~L.~Feng, A.~Rajaraman and F.~Takayama,
  Phys.\ Rev.\ D {\bf 68} (2003) 063504
  [hep-ph/0306024].
   
 \bibitem{Ade:2013zuv}
  P.~A.~R.~Ade {\it et al.} [Planck Collaboration],
  Astron.\ Astrophys.\  {\bf 571} (2014) A16
  [arXiv:1303.5076 [astro-ph.CO]].
 
 \bibitem{Fong:2013wr}
  C.~S.~Fong, E.~Nardi and A.~Riotto,
  Adv.\ High Energy Phys.\  {\bf 2012} (2012) 158303
  [arXiv:1301.3062 [hep-ph]].
  
  \bibitem{Hasenkamp:2010if}
  J.~Hasenkamp and J.~Kersten,
  Phys.\ Rev.\ D {\bf 82} (2010) 115029
  [arXiv:1008.1740 [hep-ph]].

\bibitem{Baltz:2001rq}
  E.~A.~Baltz and H.~Murayama,
  JHEP {\bf 0305} (2003) 067
  [astro-ph/0108172].



\bibitem{LMN}
T.~Li, H.~Murayama and S.~Nandi, unpublished.

\bibitem{Jiang:2006hf} 
  J.~Jiang, T.~Li and D.~V.~Nanopoulos,
  Nucl.\ Phys.\ B {\bf 772}, 49 (2007)
  [hep-ph/0610054].

\bibitem{Barger:2006fm} 
  V.~Barger, J.~Jiang, P.~Langacker and T.~Li,
  Int.\ J.\ Mod.\ Phys.\ A {\bf 22}, 6203 (2007)
  [hep-ph/0612206].

\bibitem{Barger:2007qb} 
  V.~Barger, N.~G.~Deshpande, J.~Jiang, P.~Langacker and T.~Li,
  Nucl.\ Phys.\ B {\bf 793}, 307 (2008)
  [hep-ph/0701136].


\bibitem{Jiang:2008yf} 
  J.~Jiang, T.~Li, D.~V.~Nanopoulos and D.~Xie,
  Phys.\ Lett.\ B {\bf 677}, 322 (2009)
  [arXiv:0811.2807 [hep-th]].


\bibitem{Jiang:2009za} 
  J.~Jiang, T.~Li, D.~V.~Nanopoulos and D.~Xie,
  Nucl.\ Phys.\ B {\bf 830}, 195 (2010)
  [arXiv:0905.3394 [hep-th]].

\bibitem{Li:2009cy} 
  T.~Li,
  Phys.\ Rev.\ D {\bf 81}, 065018 (2010)
  [arXiv:0905.4563 [hep-th]].
  
\bibitem{Calibbi:2009cp}
  L.~Calibbi, L.~Ferretti, A.~Romanino and R.~Ziegler,
  Phys.\ Lett.\ B {\bf 672} (2009) 152
  [arXiv:0812.0342 [hep-ph]].




  

\end{thebibliography}
\end{document}